\documentclass[11pt,a4paper]{article}
\usepackage{form}
\usepackage{overcite}
\begin{document}

%\newcommand{\mib}[1]{\mbox{\boldmath$#1$}}
%%\newcommand{\tx}[1]{\textrm{\scriptsize #1}}

%ver.1.3 since  7 April 2022
%ver.1.4 since 15 April 2022
%ver.1.5 since 18 April 2022 by ST
%ver.1.6 since 27 April 2022 by AN

%ver.2.0 since 19 September 2022 by KH; after reviewer comments
%ver.2.1 since  4 November 2022; after acceptance

\newcommand{\revone}[1]{\textcolor{black}{#1}}  
\newcommand{\revtwo}[1]{\textcolor{black}{#1}}  

%\rhead{ver.1.6 \hspace{2mm} October 20, 2021 \hspace{2mm} KH}
\rhead{}
\lfoot{{\it Preprint submitted to Reviews of Modern Plasma Physics}}
\cfoot{}
\rfoot{{\it \today}}

%\title[Waves in planetary dynamos]{Waves in planetary dynamos}

\begin{center}
{\Large Waves in planetary dynamos}\\
\vspace{3mm}  

K. Hori$^{1*}$, A. Nilsson$^2$, S.~M. Tobias$^3$\\
\vspace{3mm}

$^1$Graduate School of System Informatics, Kobe University, Rokkodai 1-1, Nada, Kobe 657-8501, Japan.  \\
$^2$Department of Geology, Lund University. S\"{o}lvegatan 12, Lund 22362, Sweden.\\
$^3$Department of Applied Mathematics, University of Leeds, Woodhouse Lane, Leeds LS2 9JT, UK.\\
$^*$The corresponding author. 
\end{center}

\renewcommand{\abstractname}{}
\vspace{-10mm}

%%%%% ABSTRACT %%%%%%%%%%%%%%%%%%%%%%%%%%%
%\noindent\rule[.6ex]{\linewidth}{.06ex}
\begin{abstract}
%\begin{center}
%\begin{minipage}{.9\textwidth}
\noindent\rule[.6ex]{\linewidth}{.06ex}

%\linenumbers

% {\sc Abstract}:
\noindent
This Special Topic focuses on magnetohydrodynamic (MHD) processes in the deep interiors of planets,
 in which their fluid dynamos are in operation. 
The dynamo-generated, global, magnetic fields provide a background for our solar-terrestrial environment. 
Probing the processes within the dynamos is a significant theoretical and computational challenge and any window into interior dynamics greatly increases our understanding.
Such a window is provided by exploring rapid dynamics,
 particularly MHD waves about the dynamo-defined basic state. This field is the subject of current attention  
 as geophysical observations and numerical modellings advance.  
We here pay particular attention to torsional Alfv\'{e}n waves/oscillations and magnetic Rossby waves,
 which may be regarded as typical axisymmetric and nonaxisymmetric modes, respectively,
 amongst a wide variety of wave classes of rapidly-rotating MHD fluids. 
The excitation of those waves has been evidenced for the Earth --- 
 whilst their presence has also been suggested for Jupiter.
We shall overview their dynamics, summarise our current understanding, and give open questions for future perspectives. \\

\noindent
Key words: MHD, rotating fluids, waves, dynamos, planets

\noindent\rule[.6ex]{\linewidth}{.06ex}
%\end{minipage}
%\end{center}
\end{abstract}
%%%%%%%%%%%%%%%%%%%%%%%%%%%%%%%%%%%%%%%%%%%

%\vspace{5mm}

%\linenumbers

\section{Introduction} \label{sec:intro}

\subsection{Some background on magnetic field}

\begin{figure}[h]%
\centering
%
%\includegraphics[width=0.9\textwidth]{fig.eps}
%\caption{This is a widefig. This is an example of long caption this is an example of long caption  this is an example of long caption this is an example of long caption}\label{fig1}
%
\begin{tabular}{lll}
 \hspace{-4mm} (a) &  (b) \\
 \hspace{-4mm} \includegraphics[viewport=15mm 45mm 140mm 103mm,clip,width=0.5\linewidth]{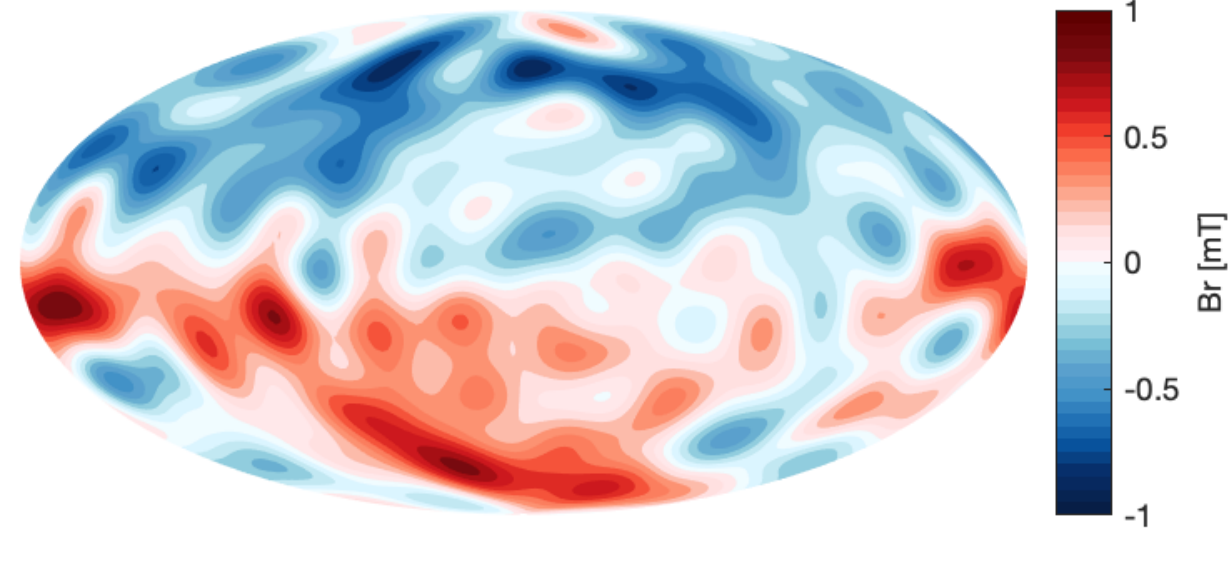} &
 \includegraphics[viewport=0mm 0mm 123mm 53mm,clip,width=0.49\linewidth]{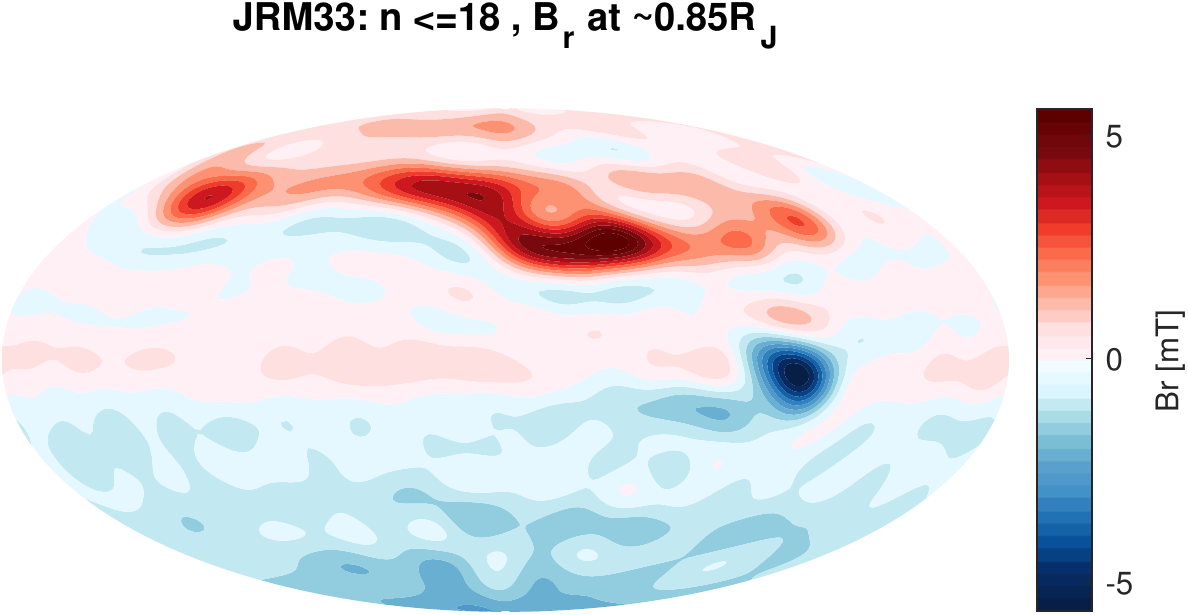} 
\end{tabular}
 \caption{(a) Earth's magnetic field in 2020 at $r = 0.546\,R_\text{E} = r_\text{core}$, the top of the fluid core 
 (\revone{reproduced from \citet{FKeta20}} with spherical harmonics of degree up to 13). %chaos7 
 (b) Jupiter's magnetic field in 2016-2021 at $r = 0.85\,R_\text{J}$, supposed to be a top of the metallic hydrogen region 
 (\revone{reproduced from \citet{Cetal21}} with spherical harmonics of degree up to 18).}  %jrm33
 \label{fig:mag}
\end{figure}

Our \revtwo{planet} has a global magnetic field
 that is predominantly an axial dipole nearly aligned with the geographical poles. 
As this field shapes part of the solar-terrestrial environment it is
of great interest in the Special Topics. 
The large-scale structure of the magnetic field, including the dipole,
 has its origin in the interior below the surface (figure~\ref{fig:mag}a). 
The field has persisted
 for at least 3.4 billion years \citep{Tetal10};
 however it is not the result of a permanent magnet --- it exhibits variations on many different timescales. For example
the dipole component has occasionally weakened and reversed its polarity on \revtwo{intervals} of the order $10^5$-$10^7$ years \citep[e.g.][]{CK95,Betal12}, whilst
 other components drift westwardly on \revtwo{periods} of the order $10^2$-$10^3$ years \citep[e.g.][]{BFGN50,NSKHH20}. Moreover 
 timeseries of the field observations repeatedly experience abrupt changes, called jerks,
 \revone{on intervals} of the order $10^0$-$10^1$ years \citep[e.g.][]{ClM84,Metal10}. 
All of these variations with an internal origin are referred to as geomagnetic secular variation.

Dynamo action is believed to operate in the interior region termed the fluid outer core,
 which is located below $\sim 54.6\%$ of the planet's radius \revtwo{$R_\text{E}$}  
 (hereafter we denote $r_\text{core} \sim 0.546 R_\text{E}$). The fluid outer core 
is made of liquid iron and some lighter chemical components. 
The origin of the dynamo process lies in the motion of the electrically conducting fluid, this induces an electric current within the region, and that maintains a magnetic field.
The detailed processes have not entirely been solved, owing to their theoretical and computational complexity; the fluid dynamics therein is likely dominated by the planet's rotation and magnetic field,  rather than inertia and viscosity (as we see below). 
Geodynamo theory therefore necessitates the understanding of the MHD of rapidly rotating fluids.
In that respect the research area shares a lot with the atmospheric and oceanic dynamics, 
 and so may be regarded as part of \revone{``}geophysical fluid dynamics''.

Some other planets are found to possess global magnetic fields
that are also thought to be generated through dynamo mechanisms (see reviews, e.g., by \citet{S10,J11,SS11} and references therein).
It is however still unclear where in those planets dynamos operate; 
 the internal structures of planets other than Earth are not well determined.
Jupiter, for example, is a gaseous planet mostly made of hydrogen and helium 
 and has the strongest planetary magnetic field (figure~\ref{fig:mag}b).
The gas giant's dynamo is likely active in the metallic hydrogen envelope; though the exact location is uncertain.  
Indeed the NASA Juno spacecraft \citep[e.g.][]{Betal17,S20} has been orbiting the planet to determine the internal structure,
 and is producing evidences that
 the conductive region likely spans up to $\sim$80-90\% of the planet's nominal radius $R_\text{J}$.
Recall that Earth's dynamo sits deep inside 
 and is masked by the rocky mantle, which acts as an insulator, screening the small-scale structure of the magnetic field.
Exploring the fields of other planets, particularly Jupiter where the conducting region is not screened as effectively, 
 could provide us with deeper knowledge about the operation of natural dynamos \citep{JH17}.

%%%%%% Figure 1 %%%%%%%%%%%%%%%%%%%%%%%%%%%%%%%%%%%%%%%%%%
\bigskip
%%%%%%%%%%%%%%%%%%%%%%%%%%%%%%%%%%%%%%%%%%%%%%%%

\quad

\lfoot{}
\cfoot{\thepage}
\rfoot{}

\subsection{A brief introduction to dynamo theory} \label{subsec:dynamo}

The early foundations of dynamo theory were made over a century ago and progress involved advances in applied mathematics and fluid dynamics. 
Early outcomes involved the demonstration of  multiple anti-dynamo theorems that limit the structure of dynamo-generated magnetic fields and the flows that generate them; 
for example Cowling's Theorem states that a purely axisymmetric magnetic field cannot be generated by dynamo action, whilst Zel'dovich's theorem maintains that
a purely two dimensional fluid motion cannot drive a dynamo;
 we refer to \citet{M78,R94,DS07,T21} for reviews. 
Those early works unraveled key elements for dynamo action, 
 one of which is the necessity of interactions between toroidal and poloidal components of the magnetic field, 
 which are defined from a decomposition of the field based on the solenoidal condition. 
\revtwo{Mean-field theory,
 where the interacting terms, e.g. the electromotive force arising from turbulent interactions are modelled or parameterised, 
 yields steady or oscillatory solutions, dependent on the relative strength of interactions;
  there is yet a wealth of theory describing limitations on the form and applicability of the interaction terms \citep{M78,HY08,T21}}.
This theory yields the basics of how a global magnetic field can be maintained and also mechanisms for periodic cycles.

In the age of computational physics and geophysics, numerical investigations have been pursued to solve the self-consistent dynamo problem, where the magnetic field is destabilised and sustained by fluid motions
 that are driven, for example, by buoyancy, and acts back on the flows that are driving it. 
Buoyancy-driven convection is thought to be a primary source of planetary dynamos,
 including Earth and Jupiter, where the planet's thermal evolution is active. Convective instability has been greatly studied, alongside the dynamo instability and 
 we refer to \citet{J15} for reviews. 
Numerical dynamos driven by convection in spherical shells first succeeded in the 1990s 
 in reproducing the generation of global magnetic field and its polarity reversals \citep{GR95,KS95}. 
This was followed by many numerical simulations (see reviews by \citet{CW15,J11})
 to reveal their scaling properties \citep[e.g.][]{CA06,D13}, 
 and to replicate individual planets and moons including  gaseous planets \citep{J14,GWDHB14}. Owing to the large separation of timescales that are relevant to the dynamics of Earth's magnetic field,
early geodynamo simulations suffered from an issue such that 
 their self-generated magnetic fields did not represent the physical regime expected to pertain to Earth's interior,
 called the magnetostrophic regime (see below).
However \revtwo{recently a consensus is building that simulations are beginning to enter the relevant 
regime \citep[e.g.][]{YGCDR16,D16,SJNF17,AGF17};
there are still ongoing debates, for example, on the existence of strong-/weak- field branches \citep{D16} and on the lengthscale-/spatial- dependence of the force balance \citep{SGA19}, 
i.e. on what scales the viscous balances pertain.}

\quad

\subsection{Detailed observations} \label{subsec:obs}

Concurrently, geophysical observations together with data modelling techniques have been advancing enormously.
Ground-based measurements and archaeological/sedimentary records
 have been analysed to enable the recovery of the field evolution for past few thousand years down to a spatial wavelength of ~72$^\circ$ \citep[e.g.][]{KCDH11,NHKSH14,HG18}.
This has enabled the description, for example, of millennial-scale high-latitude westward drifts \citep{NSKHH20} and spikes \citep{DC17}. Moreover, 
today's satellite missions, including the Swarm mission, have mapped in detail the variations of the present-day geomagnetic field.
Global models based on such measurements have established detailed 
 descriptions of the secular \revone{``}variation'',
 defined by the temporal derivative of the interior-origin field, 
 and also the \revone{``}acceleration'' (the second derivative)
 arising from the fluid core \citep[e.g.][]{FKeta20}. 
This analysis has led to the discovery of rapid dynamics, such as the several-year westward drift near the equator \citep{CAM15} and the polar jet \citep{LHF17}.

\begin{figure}[h]%
\centering
 \begin{tabular}{ll}
  (a) & \hspace{5mm}(b) \\
 \includegraphics[width=0.3\linewidth]{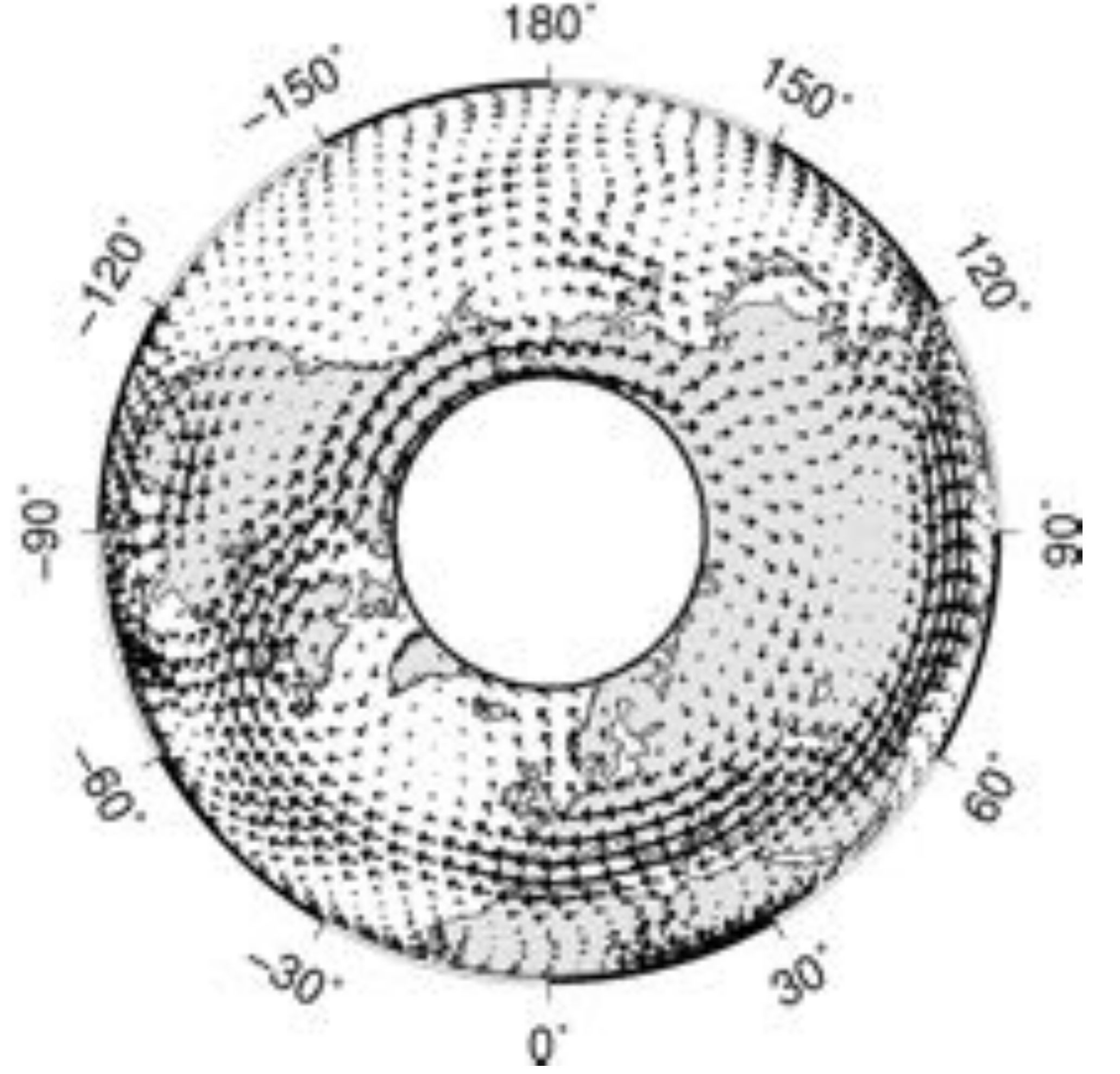} 
 & \hspace{5mm} 
 \includegraphics[width=0.32\linewidth]{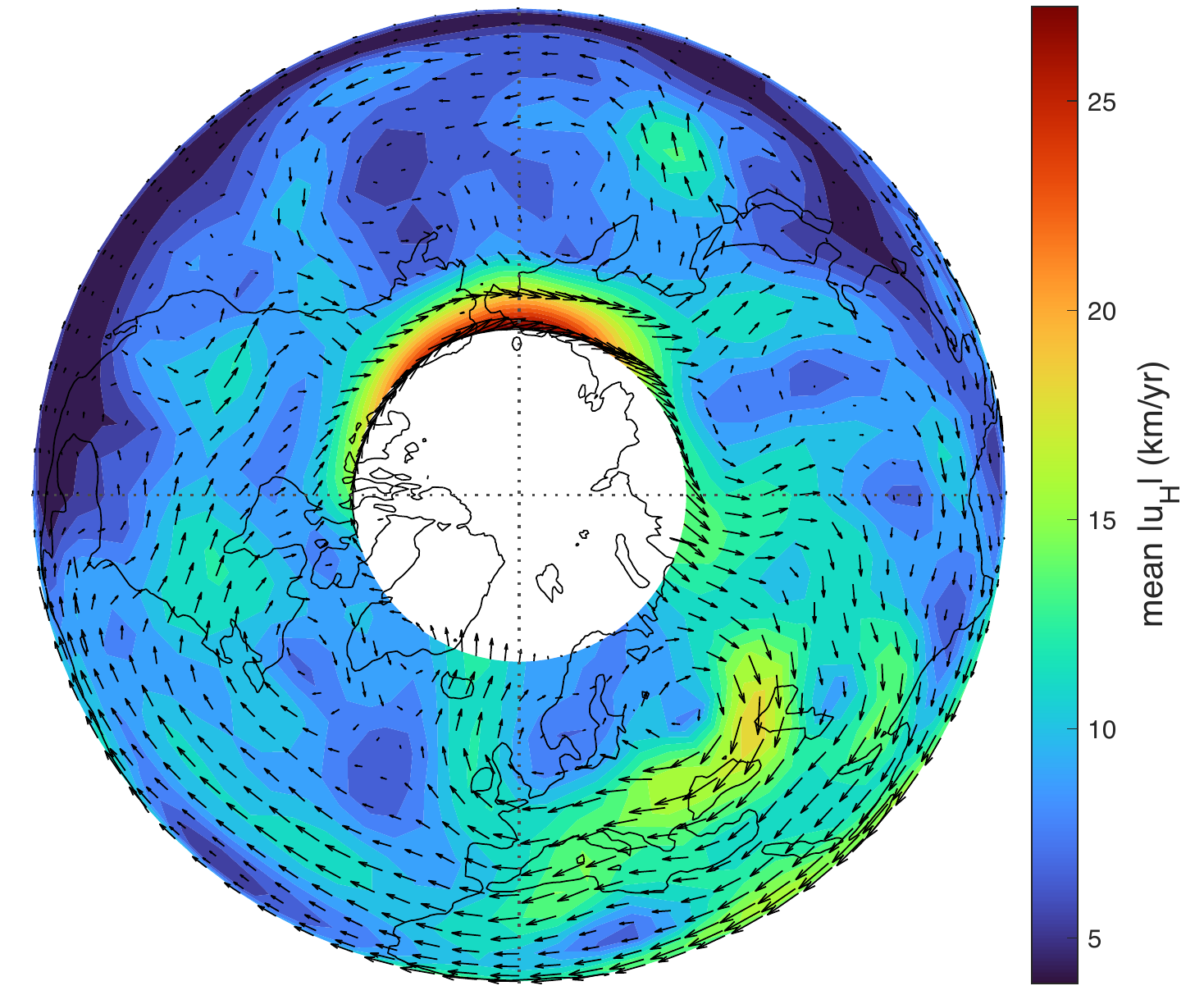} 
 \end{tabular}
 \caption{Core flows inverted from geomagnetic secular variation. 
 Arrows represent velocity, of the order $10^{-3}$ m/s, in the equatorial plane. Averages (a) over 1840-1990, adapted from \citet{PMS15}, and (b) over 1890-2010, reproduced from \citet{GHA19}.} 
 \label{fig:gyre}
\end{figure}

It has also enabled inversions to describe the fluid motion over the dynamo region:
 such an outcome is called the \revone{``}core flow model''. 
We refer to \citet{H15} for a review.
A prominent feature found by those inversions is a single anticyclonic vortex,
 sometimes referred to as the eccentric gyre (figure~\ref{fig:gyre})
 \citep[e.g.][]{PJ08}, 
 which has likely persisted for more than 100 years, although some fluctuations are observed. 
Currently data assimilation, where observations are combined with theoretical dynamo models, is becoming a common technique to provide core flow information \citep[e.g.][]{FAT11,GHA19}.
It could therefore also be quite natural to construct core flow models for other planets. This has, in part,  been attempted for Jupiter,
 in which the magnetic secular variation was realised in early missions \citep[e.g.][]{RH16};
 now the Juno mission is going to provide more details 
 \citep{Metal19,Betal22}.

%%%%%% Figure 2 %%%%%%%%%%%%%%%%%%%%%%%%%%%%%%%%%%%%%%%%%%
\bigskip
%%%%%%%%%%%%%%%%%%%%%%%%%%%%%%%%%%%%%%%%%%%%%%%%

\quad

\subsection{Waves: the focus of the Special Topics} \label{subsec:wave}

With motivation from the advances described above, we here focus on MHD waves in planetary dynamos. 
They seemingly give a framework for thinking about the secular variation/acceleration and rapid dynamics \revtwo{(with a timescale of hundreds of years),
 in contrast to convection or dynamo action, 
 in which diffusion matters and for which a typical timescale is of the order of $10^5$ years and more in Earth, for example.}
Indeed, exploring waves is highly beneficial  
 as their properties could yield information about the interior that is inaccessible %directly.
 through direct observations.
The related science of seismology has successfully scanned the elastic structure of our terrestrial planet,
 however is hindered from probing the fluid dynamo region in detail. 
An alternative investigation, utlising MHD waves could potentially visualise this, sensing the hydromagnetic properties
 such as the poloidal and toroidal component of the magnetic field; in some sense there is an analogy with the study of the properties of the solar atmosphere using \revone{``}coronal seismology''.

%\quad

We give a mathematical formulation to describe our problem below. 
\revtwo{We begin with the fluid (MHD) description. 
We here consider anelastic fluids where the Lantz-Braginsky-Roberts-Jones formalism \citep[e.g.][]{BR95,JKM09,Jetal11} is adopted, which enables the inclusion of stratification for subsonic flows.}
We assume that
 the equilibrium state is close to adiabatic, well-mixed, and hydrostatic with density $\overline{\rho}$.
The velocity of interest \mib{u} is subsonic
 (cf. the sound/seismic waves of the order $10^4$\,m/s in Earth) 
 so that the continuity equation becomes
 \begin{equation}
  \nabla \cdot \overline{\rho} \mib{u} = 0  \; .\label{eq:continuity}
 \end{equation}
\revtwo{For simplicity, we assume that the basic state $\overline{\rho}$ depends solely on spherical radius, $r$.}
Focusing on the dynamics 
 \revtwo{whose characteristic timescales are shorter than the diffusion times (see above),}
 we then consider the momentum equation
\begin{equation}
  \overline{\rho} \left[ \frac{\partial \mib{u}}{\partial t} + (\mib{u} \cdot \nabla ) \mib{u}
 + 2 \mib{\Omega} \times \mib{u} \right]
  = -\nabla p^\prime %\hat{p}
   + \mib{j} \times \mib{B}
  \; ,    \label{eq:momentum}
\end{equation}
 \revtwo{where $\mib{\Omega}$ is} the rotational angular velocity, 
 $\mib{j}$ is the current density, $\mib{B}$ is the magnetic field,
 and $p^\prime$ is a reduced pressure incorporating
 the gravitational potential. 
The induction equation for magnetic field \mib{B} is given by  
\begin{equation}
 \frac{\partial \mib{B}}{\partial t} + \mib{u}\cdot\nabla\mib{B}
  = \mib{B}\cdot\nabla\mib{u} - (\nabla\cdot\mib{u})\mib{B}    \label{eq:induction} 
\end{equation}
where the MHD approximation has been made in the non-relativistic Maxwell equations and combined with Ohm's law for a moving condutor. Of course the magnetic field is solenoidal.
The density variation in Earth's fluid core is likely less than 20\%,
 so for models of the Earth $\overline{\rho}$ in those equations may be assumed to be constant
 and the theory reduces to the incompressible, Boussinesq equations of MHD. 
In Jupiter, by contrast, the density varies by orders of magnitude.
Wave solutions are obtained as fluctuations about 
 a basic (or background) state of density, flow, and magnetic field.
Conversely, measurement of wave properties such as frequencies
 could allow the inference of physical quantities of the background media.

\revone{
Here we note the above equations give an ideal framework to examine the wave dynamics particularly and ought to be distinguished from those for dynamo action and convection (see above). 
Current planetary dynamo simulations, as discussed in sec.~\ref{subsec:dynamo}), mostly solve equations including buoyancy, viscosity and diffusion. 
Below we shall adopt those simulations to examine to what extent our diffusion-free, buoyancy-free framework could be beneficial.}

We first consider some dimensionless parameters to define the dynamical regime of interest.
A key parameter is the Rossby number $Ro = U/L\Omega$,
 where $U$ and $L$ are a typical velocity and lengthscale respectively; this quantifies the relative strength of the inertia to the Coriolis force in the momentum equation. 
In Earth's fluid core $Ro \sim 1\times 10^{-5}$,
 provided the speed $U$ is represented
 \revone{by} the slow inverted core flows of $\sim 10^{-3}$ m/s
 and $L \sim 10^6$ m (a global scale),
 and $\Omega \sim 7.3 \times 10^{-5}$ s$^{-1}$.
Jupiter's metallic region likely has $Ro \sim 6\times 10^{-6}$
 for $U \sim 10^{-2}$ m/s, $L \sim 10^7$ m, and $\Omega \sim 1.8 \times 10^{-4}$ s$^{-1}$.
Those suggest a minor role of the inertia, compared with rotation --- at least on large lengthscales (and even for quite small scales!)\revone{.} 
For hydrodynamic flows, the fluid motion in those situations
 may be expected to be two dimensional (invariant in the direction of rotation),
via the Proudman-Taylor theorem;  
 such a mode is often called geostrophic, where the Coriolis force is largely balanced by a pressure gradient. 
In the presence of magnetic field, it is possible to have a force balance where
 the Coriolis, Lorentz, and pressure gradient forces are important; this is called a magnetostrophic balance and we stress that different balances may be found at different scales.
These balances will have a significant impact on the dynamics, not only wave dynamics but also that relating to convection and dynamo action. 
Overviewing all aspects is beyond the scope of the present paper.
Here we just comment that waves will be capable of diagnosing such a dynamo state and focus on that aspect.

\quad

Analysis of MHD waves in rotating fluids dates back to the 1950-60s \citep[e.g.][]{L54,H66,M67,Bra67,AH73}.
Those ideas were examined further, as geomagnetic modelling and core flow inversions were upgraded \citep[e.g.][]{ZB97,FJ03}.
We are in 
an exciting era, where new data and tools are increasingly arriving (sec.~\ref{subsec:obs}) and this has led to a re-invigoration of
theoretical investigations, as well as observational explorations \citep[e.g.][]{GJCF10,FDCP10,B14}. 
This is intrinsically linked to --- and sheds light on --- geophysical issues such as 
the existence of a thin stably-stratified layer atop Earth's fluid core, overlying the main dynamo region. 
Waves in such a stratified environment \revone{were} termed Magnetic-Archimedes-Coriolis (MAC) waves \citep{Bra67}. 
This contrasts with those in unstratified situations, where such waves are referred to Magnetic-Coriolis (MC) waves.

What makes the subject attractive is the wide variety of different wave classes. 
This arises from the combination of MHD, rotation, and stratification in some cases; all of which singly
are classic research areas \revone{in fluid dynamics} --- 
 however their combination yields a unique physics. 
(This could be analogous to the situation for plasma physicists,
 who seem to enjoy another blend with particles, compressibility, and so on.)
The complex situation is manifested even for linear waves where
classification itself is an ongoing topic of current research; spherical geometry and the morphology of the basic magnetic field makes the problems distinctive.
For example, a comprehensive investigation in a spherical, magnetised, shallow-water system was recently made by \citet{MJT17},
 where a simple background field was assumed;  
 here recall an equivalent analysis in the hydrodynamic case was made by \citet{LH68}.
Further works are finding peculiar eigenmodes such as equatorially or polarly trapped ones  \citep[e.g.][]{BM19,N20,CAB21}.

Below
 we overview our recent work, linking to the broader subject.
The focus here is two wave classes, 
 torsional Alfv\'{e}n waves (section~\ref{sec:TAW}) and magnetic Rossby waves (section~\ref{sec:MRW}).
They are characteristic axisymmetric and nonaxisymmetric MC modes respectively that can be excited within dynamo regions. 
Their dynamics are dictated by the magnetostrophic balance in rapidly-rotating systems. 
We examine the fundamentals of such waves
 and exemplify their role in Earth and Jupiter. 
This will give us the physical basis to consider further complications,
 such as the introduction of stratification to the waves.

\quad

\section{Torsional Alfv\'{e}n waves} \label{sec:TAW}

We initially examine an axisymmetric mode, termed a  torsional oscillation or torsional Alfv\'{e}n waves. 
We begin with some fundamentals of the theory (sec.~\ref{sec:TAW_theory}), and then examine their importance for Earth (sec.~\ref{sec:TAW_geo});
we than provide a re-examination in updated geomagnetic datasets (sec.~\ref{sec:TAW_geo_data}) and potential implications in Jupiter (sec.~\ref{sec:TAW_jup}).

\quad

\subsection{Fundamentals}  \label{sec:TAW_theory}

We here outline the basic theory, following the literature \citep[e.g.][]{Bra70,RA12,JF15,HTJ19}.
Our basic equation is the azimuthal component of (\ref{eq:momentum}) in cylindrical polar coordinates $(s, \phi, z)$ where the $z$ coordinate is supposed parallel to the rotation axis $\mib{\Omega}$. 
To seek the axisymmetric two-dimensional component, we take the integral over cylindrical surfaces
 along the rotational axis to yield 
\begin{eqnarray}
 \frac{\partial}{\partial t} \langle \overline{\rho}\,\overline{u_\phi} \rangle
 &=& - \left\langle \overline{\hat{\mib{e}}_\phi \cdot ( \nabla \cdot
         \overline{\rho} \mib{u} \mib{u}}) \right\rangle
  - 2 \Omega \langle \overline{\rho}\,\overline{u_s} \rangle 
   + \left\langle \overline{\hat{\mib{e}}_\phi \cdot \frac{1}{\mu_0}(\nabla\times\mib{B}) \times \mib{B}}\right\rangle \nonumber \\
 &\equiv & F_\text{R} + F_\text{C} + F_\text{L} \; ,   \label{eq:N-S_uphi}
\end{eqnarray}
where $\mib{j} = (\nabla \times \mib{B})/\mu_0$ is the current, the magnetic permeability $\mu_0 = 4\pi \times 10^{-7}$ in SI units, and $\hat{\mib{e}}_z$ is the unit vector in the azimuthal direction. 
Here $\overline{f}$ and $\langle f \rangle$ denote the $\phi$-average (i.e. the axisymmetric part)
 and the $z$-average from $z_{+}$ to $z_{-}$, respectively, for an arbitrary function $f$.
Outside the tangent cylinder, which is an imaginary cylinder circumscribing the inner shell,
 $z_\pm = \pm \sqrt{r_\text{o}^2 - s^2} \equiv \pm H$
 where $r_\text{o}$ is the radius of the conducting region:
 hereafter we only consider the region outside the tangent cylinder. 
From the divergence theorem, and the continuity equation (\ref{eq:continuity}),
 \revone{the Coriolis force $F_\text{C}$} vanishes, i.e. there is no net mass flux across a given cylindrical surface. 
\revone{When the inertia including the Reynolds term $F_\text{R}$ (and viscosity) is negligible compared with the Coriolis and Lorentz forces $F_\text{L}$} (i.e. magnetostrophic balance), 
 the equation yields a steady state,  
 $\int { \hat{\mib{e}}_\phi \cdot (\nabla\times\mib{B}) \times \mib{B}/\mu_0\,}dS  = 0$, 
 termed the Taylor state \citep{Tay63}
 by which the magnetic field configuration is constrained.

\begin{figure}[h]%
 \centering
 \begin{tabular}{cc}
  \includegraphics[width=0.3\linewidth]{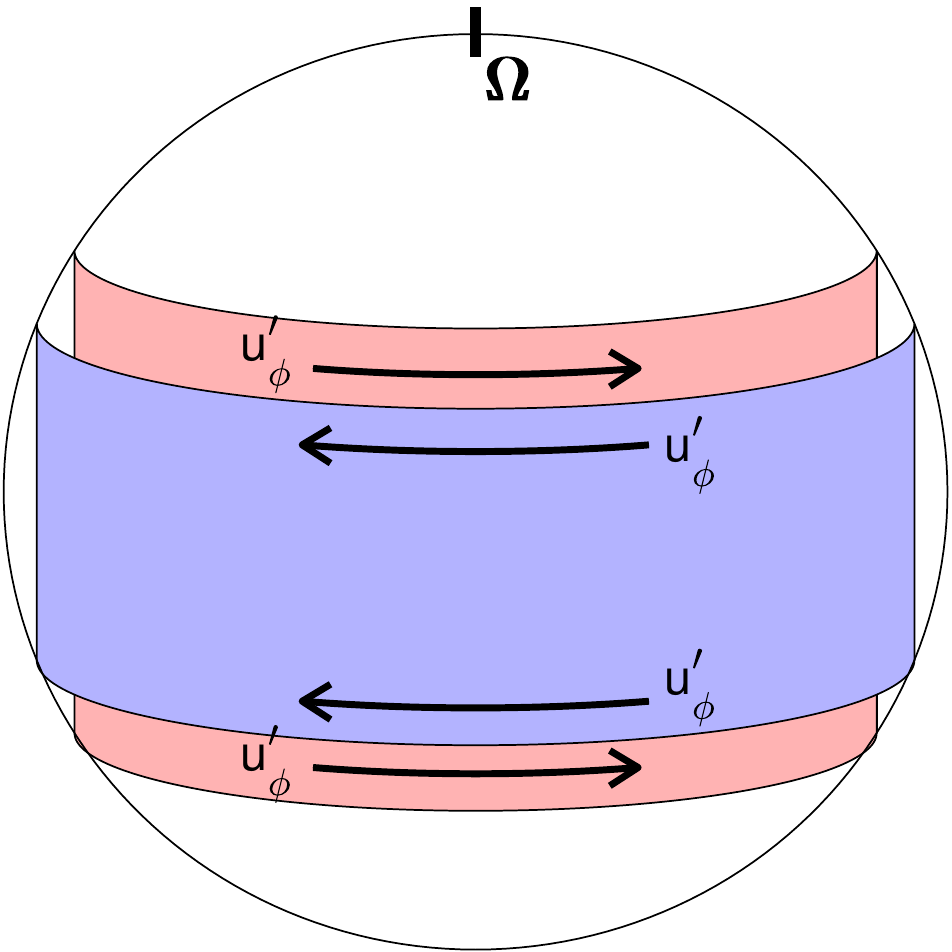}  & 
  \hspace{10mm}
  \includegraphics[width=0.3\linewidth]{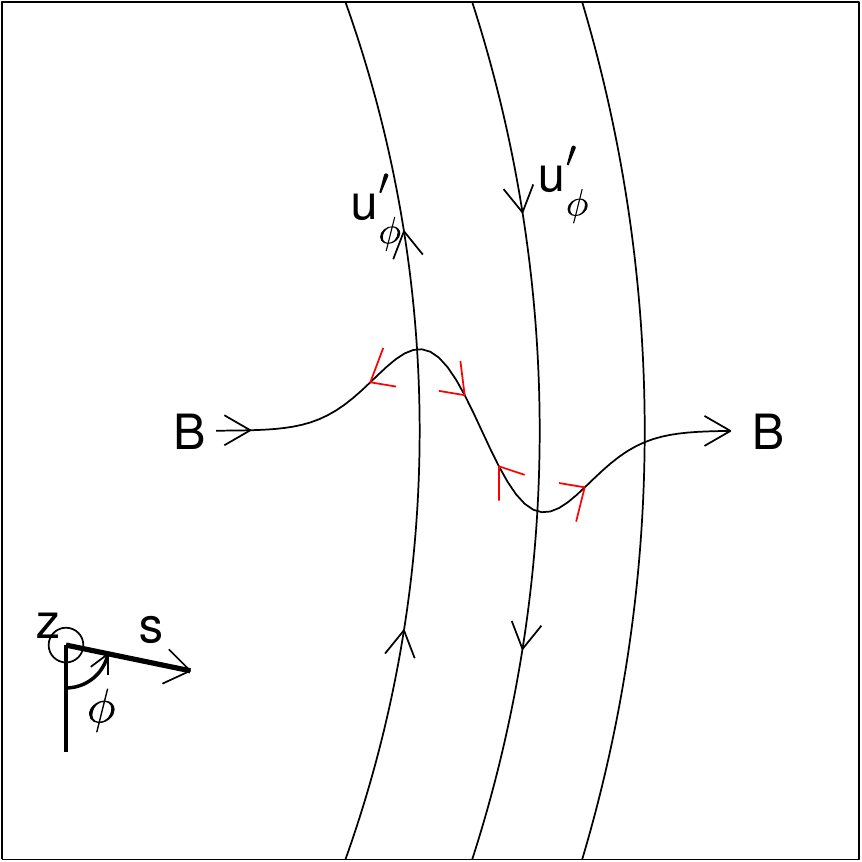} 
 \end{tabular}
 \caption{Schematic illustration of torsional Alfv\'{e}n waves/oscillations (adapted from \citet{HJAFT}).}
 \label{fig:TAW}
\end{figure}

Allowing small perturbations about this state yields waves/oscillations 
 (see the detailed derivation \revtwo{in \citet{TJT14,HTJ19}}). 
We split magnetic field and velocity into their temporal mean and fluctuating parts,
 which are hereafter denoted by tildes and primes, repectively, i.e. $\widetilde{f} = (1/\tau) \int f dt$ and $f' = f - \widetilde{f}$ where $\tau$ is a time window of integration and $\widetilde{f'} = 0$. 
Substituting the induction equation (\ref{eq:induction}) into the Lorentz term $F_\text{L}$ of (\ref{eq:N-S_uphi})
 and assuming that an ageostrophic term is sufficiently small,
 we get a single equation: 
\begin{equation}
 \frac{\partial^2}{\partial t^2} \frac{\langle \overline{u'_\phi} \rangle}{s} 
 - \frac{1}{s^3 h \langle \overline{\rho} \rangle } \frac{\partial}{\partial s}
     \left( 
       s^3 h \langle \overline{\rho} \rangle
       U_\text{A}^2
       \frac{\partial}{\partial s} \frac{\langle \overline{u'_\phi} \rangle}{s}
     \right) 
 =  \frac{\partial }{\partial t} \frac{F_\text{R} + F_\text{LD}}{s \langle \overline{\rho} \rangle } \;,
  \label{eq:TAW}
\end{equation}
where $h = z_{+} - z_{-}$ is the height of the cylinder of radius $s$ along the $z$ axis, 
 i.e. $h=2H$ outside the tangent cylinder.
The left hand side of (\ref{eq:TAW}) presents
 the homogeneous part of the PDE and the equation of torsional Alfv\'{e}n waves.
Terms on the right hand side can be interpreted as forcing to the wave equation,
 where $F_\text{LD}$ denotes the Lorentz force $F_\text{L}$ excluding the restoring part for the wave.
In the restoring force
 $U_\text{A}^2 = \langle \overline{\widetilde{B_s^2}} \rangle/\mu_0 \langle \overline{\rho} \rangle$, 
 representing the squared Alfv\'{e}n speed given by cylindrical averages of radial field $B_s$.
The homogeneous equation describes waves propagating in cylindrical radius $s$
 with the speed $U_\text{A}$. 
They may travel either inwardly ($-s$) or outwardly ($+s$), and may also superpose to give a standing wave referred to as \revone{``}oscillations''.
Their schematic illustration is shown in figure~\ref{fig:TAW}. 
Typical timescales can be interannual to decadal in Earth and Jupiter (secs.~\ref{sec:TAW_geo} and \ref{sec:TAW_jup}). 
Classically, these waves are ideally supposed to be non-dispersive; 
 in reality they could be dispersive owing to the geometry and dissipation. 
The Reynolds forcing $F_\text{R}$ likely plays a minor role in Earth's fluid core,
 as there is no exchanges with the rocky mantle and the reduced importance of inertia signified by the small $Ro$. This 
 implies that the Lorentz \revtwo{force} $F_\text{LD}$ is a major driver \citep{Bra70,TJT15}. 
In Jupiter the lack of rigid boundaries and the presence of significant zonal flows \revtwo{in the molecular envelope} is likely to
 make the Reynolds forcing $F_\text{R}$ more significant than $F_\text{LD}$ \citep{HTJ19}.

%%%% new Figure 3 %%%%%%%%%%%%%%%%%%%%%%%%%%%%%%%%%%%%%%%%%%%%
\bigskip
%%%%%%%%%%%%%%%%%%%%%%%%%%%%%%%%%%%%%%%%%%%%%%%%

\quad

\subsection{Torsional waves in Earth's core}   \label{sec:TAW_geo}

Torsional waves are best illustrated in the geodynamo
 through both observations and simulations. 
Early studies \citep[e.g.][]{Bra70} sought possible wave motion
 on a $\sim$60 year timescale, which is a relevant peak in the geomagnetic variation as we shall see below.
\citet{ZB97} extended those investigations to find fluctuations in the azimuthal component
 of core flows that were inverted from the magnetic secular variation.
With the torsional oscillation theory they attempted to infer the 1d structure of the Alfv\'en speed $U_\text{A}$
 and to estimate a  $\langle \overline{\widetilde{B_s^2}} \rangle^{1/2}$ of the order 0.1 mT within the dynamo.

Meanwhile, scaling properties of convection-driven dynamos (sec.~\ref{subsec:dynamo})
 suggested an internal field strength of the order 1 mT,
 implying a shorter timescales for torsional oscillations. 
\citet{GJCF10} explored signals of several years in core flow models
 and attributed them, not the decadal signal, to the torsional oscillations.
They also evaluated the angular momentum exchange with the rocky mantle
 to show the fluctuations were compatible with a variation in length-of-day,
 i.e. the rotation rate of the planet,
 in which a period $\sim$6 years was seen.

\begin{figure}[h]%
\centering
  \includegraphics[width=0.55\linewidth]{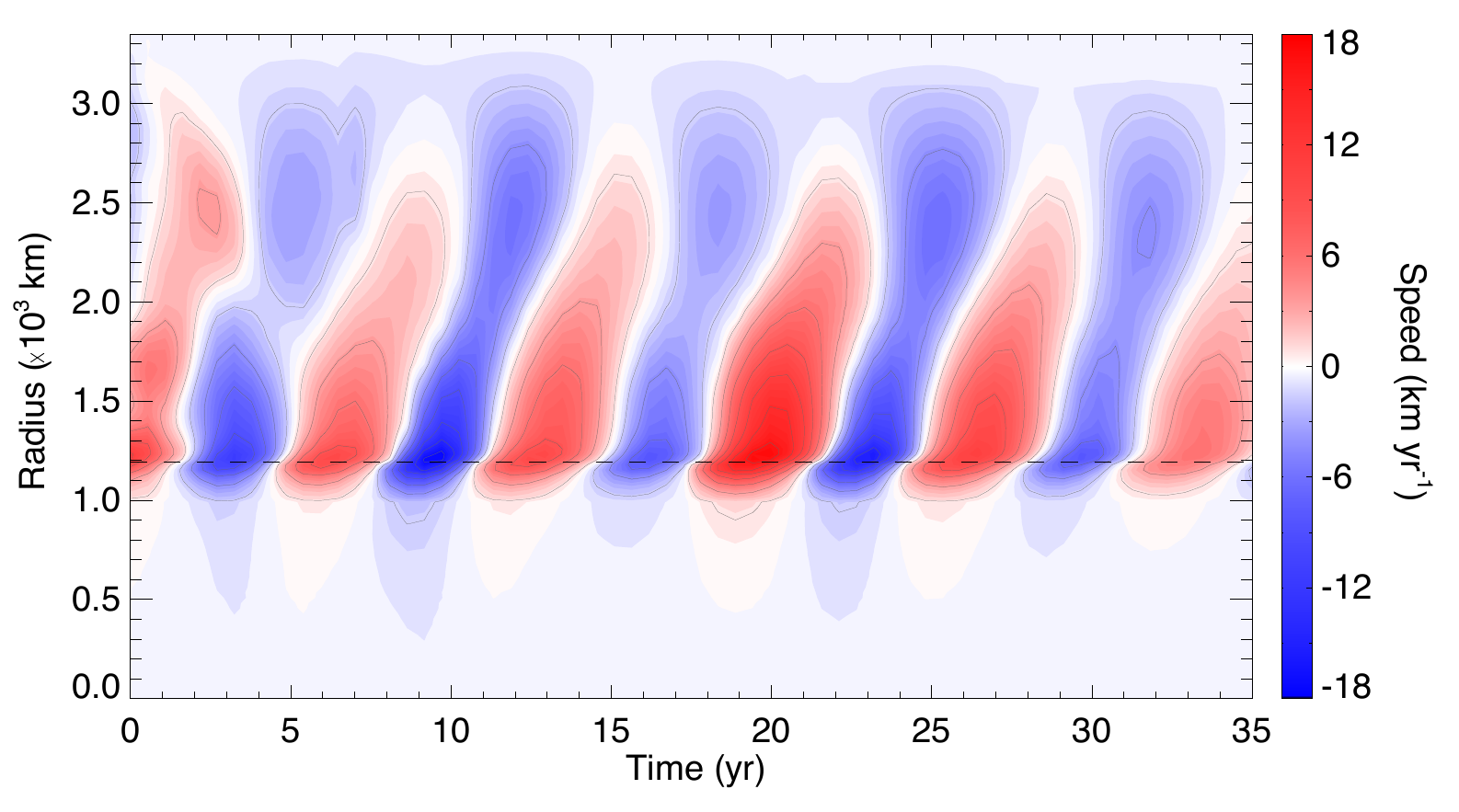} 
 \caption{Torsional Alfv\'{e}n waves seen in a spherical convection simulation for Earth's core (adapted from \citet{TJT19}). 
 The zonal velocity fluctuation $\langle \overline{u'_\phi} \rangle$ is shown in time-cylindrical radius domain.}
 \label{fig:TAW_geo}
\end{figure}

Their picture raised further interesting questions. 
First, the identified wave exhibited outward propagation toward the equator from deep
 and appeared to be excited quasi-periodically; 
\revone{when it approached the equator,} no clear reflections at the equator were observed. 
Indeed numerical dynamo simulations embraced travelling waves,
 rather than standing ones \citep[e.g.][]{WC10,TJT14,SJNF17}. 
\citet{SJ16} pointed out that the dissipation across the core-mantle-boundary would inhibit wave reflections there, to leave travelling modes only: 
 the processes were nicely demonstrated by the solution of an initial value problem \citep{GJC17}.
In contrast, spherical magnetic convection simulations with no couplings with the mantle being assumed
 reproduced the one-way propagation excited near the tangent cylinder repeatedly \citep{TJT19}: 
 figure~\ref{fig:TAW_geo} depicts such a case.  
This dynamics is likely a natural consequence of the convection in the fluid core,
 which is most vigorous near the tangent cylinder in which buoyancy sources arise from the inner core solidification. 
\revtwo{The simulation by \citet{TJT19} reveals that 
 it is possible to launch an axisymmetric disturbance of the Alfv\'{e}n frequency there,}
 which is absorbed as it approaches the rigid boundary of the rocky mantle.

However observationally, 
 whilst core flow inversions illustrate the wave-like patterns, 
 these signals do not appear in the magnetic data clearly.
\citet{SJM12} examined over decadal timeseries of the geomagnetic secular acceleration,
 $\partial^2 B_r/\partial t^2$, at chosen locations
 in terms of the Fourier transform and empirical mode decomposition
 and reported identification of $\sim$6 year periodicities. 
We shall address this in the following subsection.

%%%% new Figure 4 %%%%%%%%%%%%%%%%%%%%%%%%%%%%%%%%%%%%%%%%%%%%
\bigskip
%%%%%%%%%%%%%%%%%%%%%%%%%%%%%%%%%%%%%%%%%%%%%%%%

\quad

\subsection{Geomagnetic data revisited}   \label{sec:TAW_geo_data}

We here revisit the geomagnetic data for torsional waves, given the recent improvement in observational datasets. Moreover, data-driven techniques are drastically advancing
 and are now capable of extracting signals more efficiently. Here
we adopt a technique called dynamic mode decomposition, DMD \citep[e.g,][]{Sch10,KBBP16}. 
This may be regarded as an update of the Fourier transform and proper orthogonal decomposition (POD) --- equivalent to the principal component analysis (PCA) ---
 and may approximate spatio-temporal data in the form of the sum of normal modes\revtwo{.
For instance, a given dataset $X = X(\theta,t)$ may be approximately represented as} 
 $\sum_{j=1}^r b_j \Phi_j (\theta) \exp{(\lambda_j t)}$ 
 where $b_j$ is real and $\lambda_j$ and $\Phi_j$ are complex. 
Here $\textrm{Im}\lambda_j$, $\textrm{Re}\lambda_j$, $b_j$, and $\Phi_j (\theta)$ denote, respectively,
 the frequency, growth rate, magnitude, and spatial structure of the $j$-th DMD mode (out of the total $r$ modes).  
The outcomes can therefore be compared with normal mode solutions of the wave equation (\ref{eq:TAW})
 [see Appendix \ref{sec:normal_modes} and figure~\ref{fig:TW_normal_modes} for normal mode calculations].
The methodology was utilised in spherical MHD simulations \citep{HTT20}.

The data to be analysed is the axisymmetric fluctuating part of both the cylindrically-radial secular variation
 $\partial \overline{B_s}/\partial t$ and the azimuthal core flow $\overline{u'_\phi}$
 \revtwo{on the core surface $r_\text{core}$} between 1940-2005\revtwo{: the datasets depends on latitude $\theta$ and time $t$}.
These are computed from ensemble averages of 50 realisations given by the up-to-date assimilation model \citep{GHA19} that we refer to as cov-obs2019 hereafter.
\revtwo{The model gives the coefficients of spherical harmonics of the magnetic potential, termed the Gauss coefficients,
 and the equivalent values for the core flow. 
So we first calculate the component $\partial B_s/\partial t$ and $u_\phi$ at $r_\text{core}$, with the Schmidt normalised
associated Legendre function.
We then compute their axisymmetric parts, average over the realisations, and remove the temporal means at each $\theta$.} 
In order to seek wave signals that have the form 
 $[\overline{B'_s}, \overline{u'_\phi} ] \propto \exp{\mathrm{i} \omega t}$, 
 we put the two sets of the latitude-time data together into the decomposition analysis and perform the DMD over the dataset. 
Here we examine the data at 140 gridpoints between $\pm 69.5^\circ$ and for 65 snapshots sampled every year.
We then introduce a delay coordinate 
 to stack the data and to capture either travelling or standing waveforms \citep[e.g.][]{KBBP16}\revtwo{: the methodology is also described in \citep{HJAFT}}.
As the rms error between the input and reconstructed datasets are found to be minimised for a delay coordinate of 2, we set this parameter below. 

%\quad

\begin{figure}[h]%
\centering
\begin{tabular}{ll}
 \hspace{3mm}\includegraphics[height=0.53\linewidth]{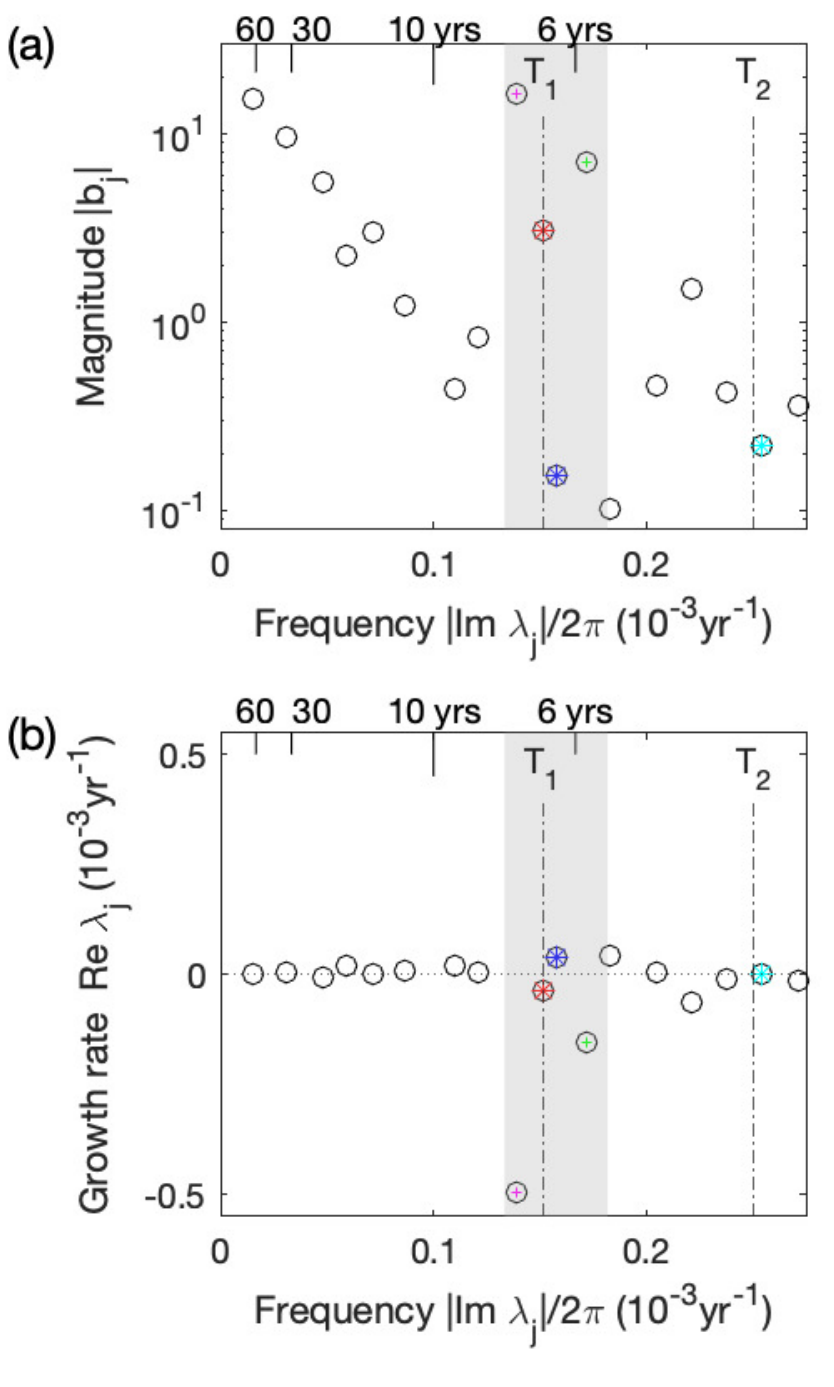} & 
 \hspace{-2mm}\vspace{-3mm}\includegraphics[height=0.5\linewidth]{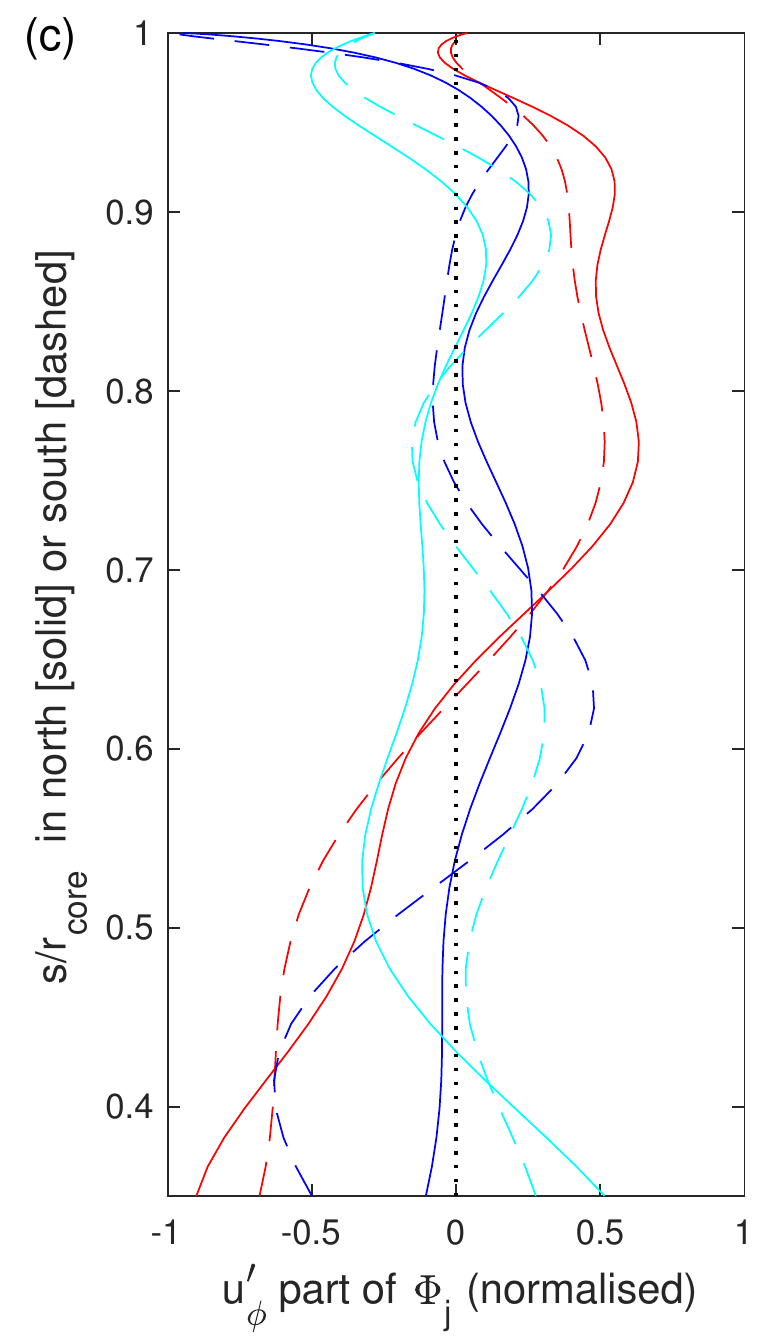} \\
 \quad \\
 \includegraphics[bb=0mm 0mm 145mm 130mm, clip, width=0.5\linewidth]{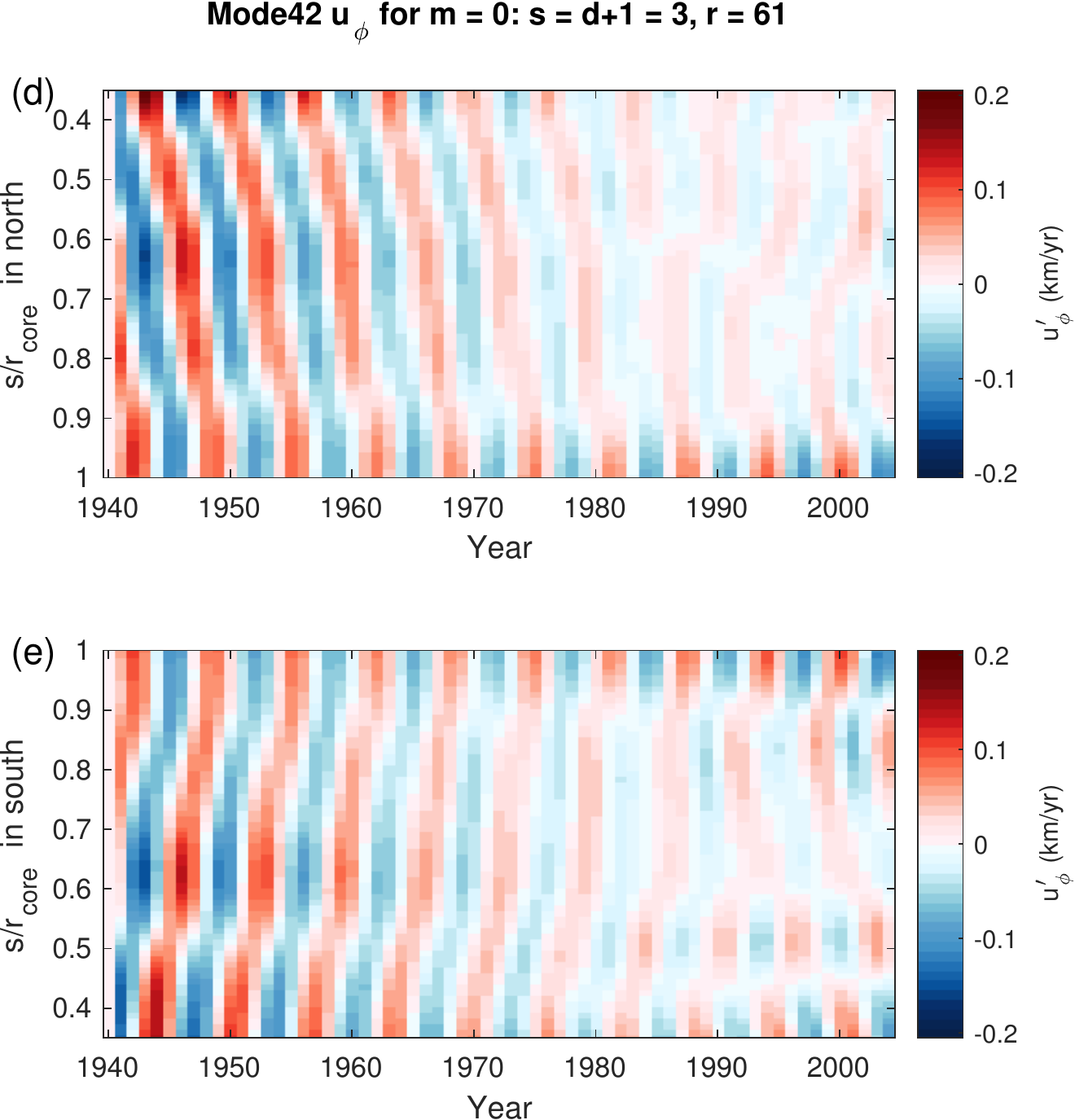} &
 \includegraphics[bb=0mm 0mm 145mm 130mm, clip, width=0.5\linewidth]{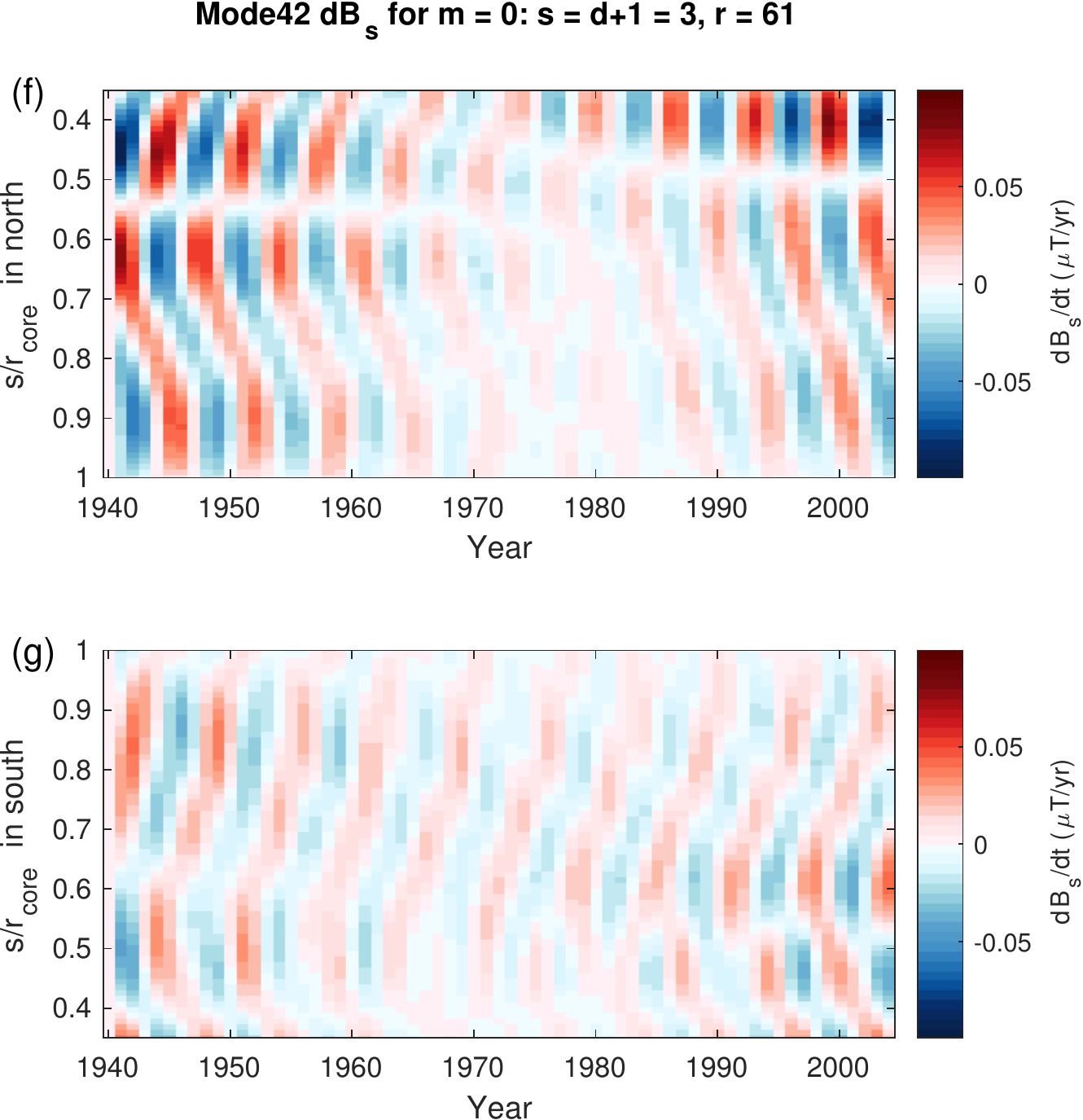} %\\
\end{tabular}
 \caption{DMD analysis of axisymmetric geomagnetic secular variation and core flow in 1940-2005
 (\revone{produced from} \citet{GHA19}).  %cov-obs2019 
 (a) Spectral and (b) dispersion diagrams of the dataset comprising of $\overline{u'_\phi}(\theta,t)$
 and $\partial\overline{B_s}/\partial t(\theta,t)$.
 Periods are represented in years on the top of each panel. 
 Symbols highlighted in color indicate Modes in a window of period 5.5-7.5 years (shaded region). 
 Individual Modes in the window are highlighted by different colors and symbols:  we refer to the red and blue asterisks as Modes 1 and 2, respectively, whilst the \revtwo{magenta} and green crosses represent dissipative modes.  
 The vertical dashed-dotted line labelled by $T_i$ indicates the frequency of the $i$-th TW normal mode for a background field ${\langle \overline{\widetilde{B_s^2}} \rangle^{1/2} }\lesssim$ 3.9 mT (see fig.~\ref{fig:TW_normal_modes}a). 
 One Mode found in the vicinity of the $T_2$ line is also indicated in cyan and is referred to as Mode 3. 
 (c) Latitudinal structures of $\overline{u'_\phi}$ for Modes 1 (red), 2 (blue), and 3 (cyan) are represented with respect to $s/r_\text{core}$. 
 Solid (dashed) curves show its profile in the northern (southern) hemisphere. 
 (d-e) Reconstructed spatiotemporal structure of $\overline{u'_\phi}$ for the superposition of Modes 1-3.
 (f-g) Similar to figures~d-e but of $\partial \overline{B_s}/\partial t$. 
  In (d,f) northern and (e,g) southern hemispheres.}
  \label{fig:TAW_geo_DMD}
\end{figure}

%
Figures~\ref{fig:TAW_geo_DMD}a and b show the spectrum and the dispersion of the DMD signals,
 respectively.  
They show spectral peaks at low frequencies (corresponding to periods of 64.9, 32.3, and 20.9 years)
 and also local peaks around a period of $\sim$6 years (the shaded region).
The latter comprises of four individual modes (highlighted by coloured symbols), 
 out of which two are found to be low quality ($\text{Im}\lambda/2\text{Re}\lambda < 4$) --- in this case
 meaning highly dissipative --- 
 and two to be high quality ($\text{Im}\lambda/2\text{Re}\lambda \sim 13$).
The two wave-like modes have periods of 6.6 and 6.3 years:
 we refer to them as Mode 1 and 2 and highlight them in red and blue.  
Their latitudinal structures  with respect to $s$ are given in figure c: 
Mode 1 in red has one zero crossing at $s/r_\text{core}\sim 0.65$.
This could be a signature of the first eigenmode of torsional wave (\ref{eq:TAW})
 [see figure~\ref{fig:TW_normal_modes}a for a background $\overline{\widetilde{B_s}}$ of maximum 3.9 mT]. 
If this field structure is assumed, the second eigenmode could be predicted: 
 whose frequency is indicated by a vertical line labelled as $T_2$ in the figures a and b:
 we refer to the chosen mode as Mode 3 (in cyan). 
The three Modes 
 are reconstructed for the spatio-temporal structure of $\overline{u'_\phi}$ in figures d-e.
The pattern shows the travelling nature in either hemisphere nicely, 
 suggesting the DMD analysis reproduces the early reports \citep[e.g.][]{GJCF10}
 and extracts the relevant modes. 

The corresponding pattern in the magnetic field $\partial \overline{B_s}/\partial t$
 is now visualised in figures f-g.
We can detect some travelling features;
 however the detected signal only exhibits a magnitude of 1\% or smaller of the overall variation.
This in part explains why the torsional waves are not easily detected in magnetic data.
Meanwhile the analysis here demonstrates how the data analysis is capable of pulling out
 such a tiny, but physically important, signal.

%%%% new Figure 5 %%%%%%%%%%%%%%%%%%%%%%%%%%%%%%%%%%%%%%%%%%%%
\bigskip
%%%%%%%%%%%%%%%%%%%%%%%%%%%%%%%%%%%%%%%%%%%%%%%%

\clearpage

%%%% new Figure 6 %%%%%%%%%%%%%%%%%%%%%%%%%%%%%%%%%%%%%%%%%%%%
\bigskip
\begin{figure}[h]%
\centering
  \includegraphics[width=0.55\linewidth]{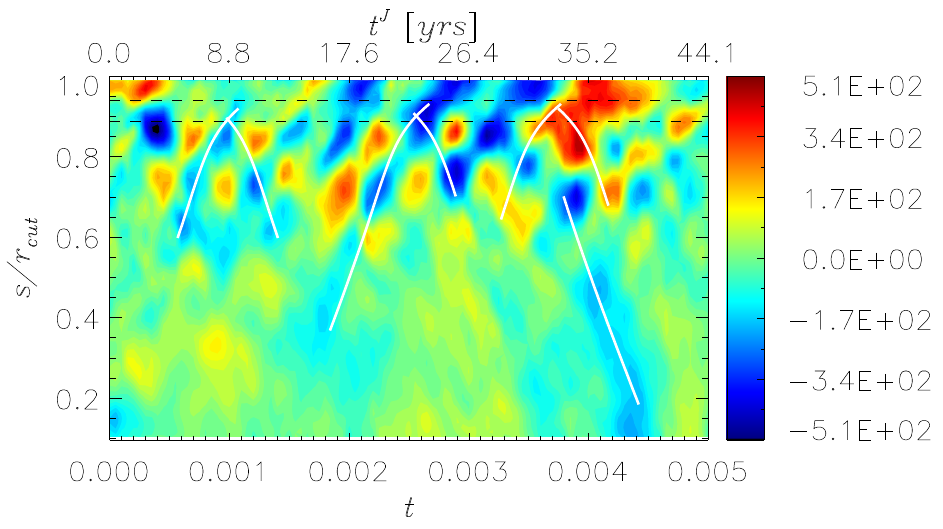}
 \caption{Similar to Fig.~\ref{fig:TAW_geo} but in a dynamo simulation
 for Jupiter's metallic hydrogen region (adapted from \citet{HTJ19}). 
 Here the cylindrical radius $s$ is normalised by the cutoff radius, $r_\text{cut} \sim 0.96 R_\text{J}$, of the simulation. White curves indicate phase paths of the Alfv\'{e}n speed $U_\text{A}$. 
 A dimensional time is represented in years on the top of the panel.
 Horizontal dashed lines indicate radii across which density and electrical conductivity drop by orders.}
 \label{fig:TAW_jup}
\end{figure}
%%%%%%%%%%%%%%%%%%%%%%%%%%%%%%%%%%%%%%%%%%%%%%%%

\quad

\subsection{Torsional oscillations in Jupiter}  \label{sec:TAW_jup}

Given the presence of torsional waves in Earth's fluid core, one might expect to find them in 
other planets. 
Indeed the potential for their discovery has been growing as planetary exploration and numerical modelling advance. 
Modern numerical dynamos, that
 implement the transition from the metallic to molecular hydrogen envelopes of Jupiter, 
 have succeeded in reproducing the dipole-dominated, global magnetic field \revtwo{\citep{J14,GWDHB14}}. 
Using those numerical models  
 \citet{HTJ19} proposed that torsional Alfv\'{e}n waves on timescale of the order 1-10 years
 could be excited in the gas giant too;
 zonal fluctuations seen for a fiducial case are exhibited in figure~\ref{fig:TAW_jup}.
Moreover the simulations demonstrated that Jovian torsional waves could be standing waves; here  
waves would partially reflect from the interface that is created by the abrupt change
 in the electrical conductivity as the metallic hydrogen transits to the molecular hydrogen. 
The ratio of reflection and transmission is essentially determined by the wavenumber of the oscillation and the skin depth for the mode.

Such fluctuations in zonal velocity could impact on the dynamics beyond the metallic region. 
One consequence could be fluctuations in length of day --- 
 as happens in Earth. 
The simulations above suggested a magnitude of 
the order $10^{-2}$ s or smaller,  so likely very tiny.
\makeatletter
Nonetheless it is worth noticing that \revtwo{the planet's rotation rate, \revone{or} the System \@Roman{3} \revone{coordinate system} determined from its periodic radio emission,} is measured to such precision and its variation was the subject of some debate \citep[e.g.][]{HCR96}. 
\makeatother
Another consequence is remarkable in a gaseous planet:
 zonal flow fluctuations arising from MHD waves may partly transmit into the overlying poorly-conducting envelope, whilst dissipating. 
This implies the deep oscillation might be probed through near-surface observations,
 such as visible, infrared, and microwave measurements.
Now the Juno's multiple measurements were reported to be consistent with
 the surface zonal wind extending down to thousands of kilometres,
 0.93-0.96$R_\text{J}$ \citep[e.g.][]{KGHetal18,Metal19}.

Interestingly, ground-based telescope observations have witnessed
 intradecadal to decadal variations of the surface \citep[e.g.][]{F17}.  
To seek the tropospheric dynamics,
 \citet{AFetal19} investigated infrared images taken at $\sim$5 $\mu$m wavelength for more than 30 years
 and found cycles of 4-9 years in latitudinal bands between \revtwo{$\sim$40$^\circ$N and $\sim$40$^\circ$S}.
Those observations might be accounted for by the torsional oscillations arising from the interior \citep{HJAFT}. 
\revtwo{The scenario necessitates the processes coupling amongst the modulations in zonal flows, the tropospheric convection, and its infrared observation.
This exemplifies the notable dynamics of a gaseous planet} in contrast with a terrestrial planet.

\quad

\section{Magnetic Rossby waves}  \label{sec:MRW}

Now we move onto nonaxisymmetric modes. 
They can be classified into three categories: Rossby waves, Alfv\'{e}n waves, and waves that have certain characteristics of both, termed magnetic Rossby waves.  
First we discuss their fundamental properties via linear theory (sec.~\ref{sec:MRW_theory})
 and explore their relevance in the geodynamo and the geomagnetic westward drift (sec.~\ref{sec:MRW_geo}); then we move on to discuss 
weakly nonlinear effects (sec.~\ref{sec:MRW_nonlinear}).

\quad
 
\subsection{Fundamentals}  \label{sec:MRW_theory}

\begin{figure}[b]%
\centering
 \begin{tabular}{ll}
  (a) & \hspace{5mm}(b) \\
  \includegraphics[width=0.4\linewidth]{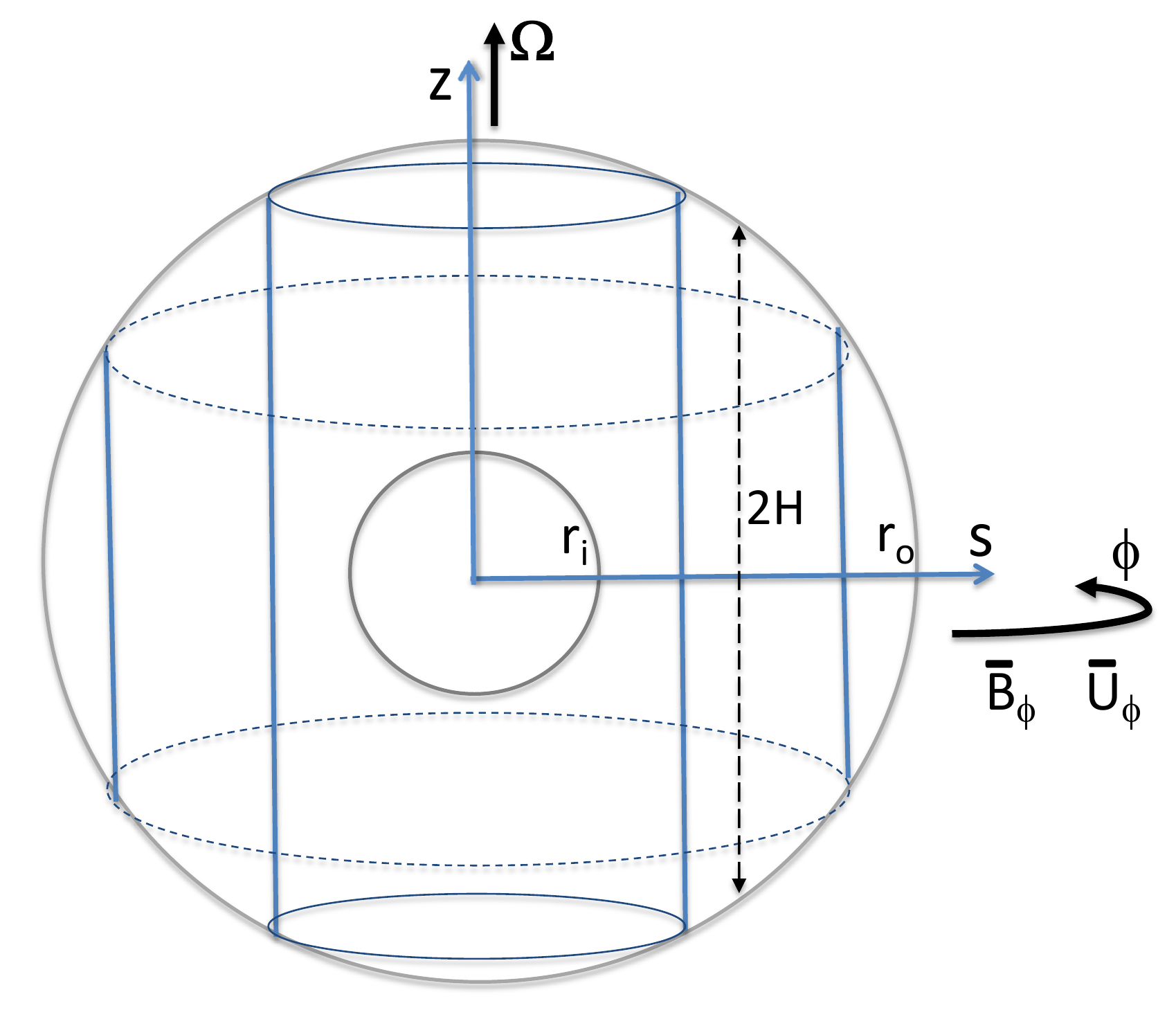} &
  \hspace{5mm}\includegraphics[viewport=5mm 15mm 85mm 75mm,clip,width=0.4\linewidth]{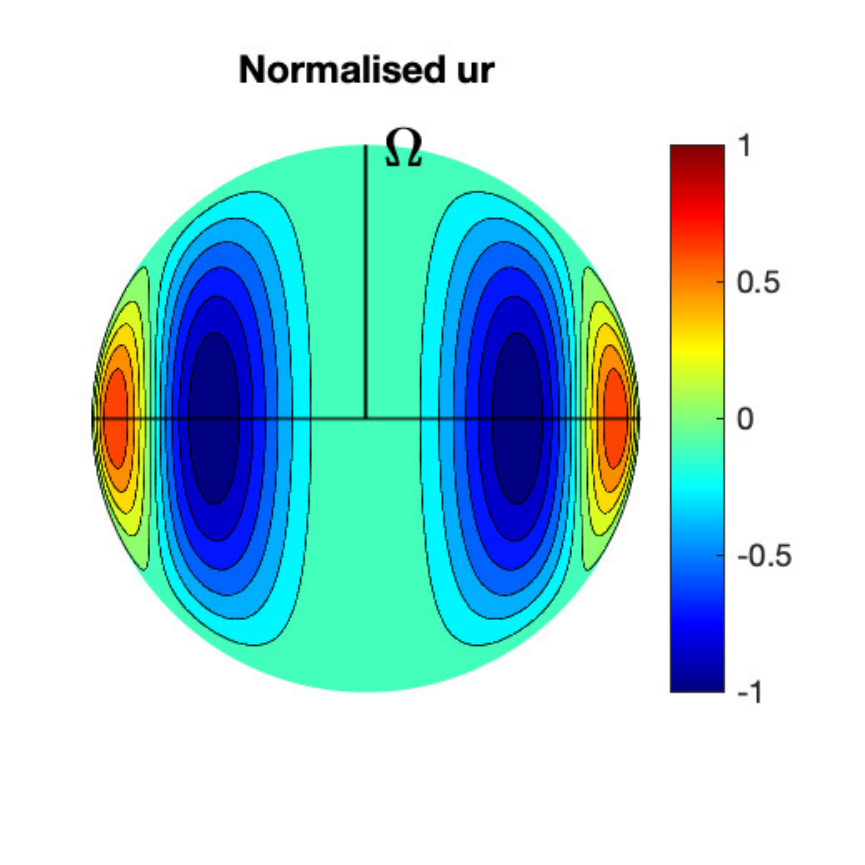} 
 \end{tabular}
 \caption{(a) The 2d quasi-geostrophic model adopted. 
 (b) An eigenfunction for background $\widetilde{B_\phi} \propto s$ and constant $\rho$ \citep{M67}.
 Normalised radial velocity $u_r$ is presented in the meridional plane with respect to the rotation axis $\mib{\Omega}$. }
 \label{fig:MRW}
\end{figure}

Guided by the literature \citep[e.g.][]{H66,HTJ18}, 
 we describe these non-axisymmetric waves for anelastic fluids. 
We here adopt an illustrative quasi-geostrophic model for rotating spherical shells \citep[e.g.][]{Bus70,Bus76,CFF14}: 
 the 2d model is schematically illustrated in figure~\ref{fig:MRW}a.
This approach ought to be distinguished from the full problem
 (\ref{eq:continuity})-(\ref{eq:induction}), as pioneered by \citet{M67}.
One of his solutions is exhibited in figure~\ref{fig:MRW}b, 
 representing a symmetric mode with respect to the equator, 
 i.e. a magnetic Rossby mode.

%%%% new Figure 7 %%%%%%%%%%%%%%%%%%%%%%%%%%%%%%%%%%%%%%%%%%%%
\bigskip
%%%%%%%%%%%%%%%%%%%%%%%%%%%%%%%%%%%%%%%%%%%%%%%%

%\quad
%
Our basic equation here is the vorticity equation. 
We take the curl of the momentum equation (\ref{eq:momentum}) in cylindrical coordinates
 to yield the equation for the axial ($z$) component of vorticity, $\mib{\xi} = \nabla\times\mib{v}$. 
This ensures conservation of the potential vorticity $(2\Omega + \xi_z)/\overline{\rho} H$
 in the absence of terms other than the inertial and Coriolis forces.
The classic Rossby waves arise from vortex tube stretching and shrinking
 to conserve the potential vorticity. 

To seek the MHD equivalent
 we consider \revtwo{the $z$-component of} the vorticity equation averaged over $z$, 
\begin{eqnarray}
 \small
\frac{\partial}{\partial t} \langle \overline{\rho} \xi_z \rangle
  + \langle \nabla_\text{H} \cdot \overline{\rho} (\mib{u} \xi_z - \mib{\xi} u_z)\rangle
%  - \langle ( \mib{u} \xi_z  - \mib{\xi} u_z  ) \cdot\nabla \overline{\rho}  \rangle  
  -2 \Omega \left\langle \frac{\partial}{\partial z} \overline{\rho} u_z
                        + u_s \frac{\partial \overline{\rho}}{\partial s} \right\rangle   
 =   \langle \nabla_\text{H} \cdot (\mib{B} J_z - \mib{J} B_z)\rangle  
 \quad 
 \label{eq:z-mean_vorticity_eq} 
\end{eqnarray}
 where $\nabla_\text{H} \cdot \mib{A} = (1/s) \partial (sA_s)/\partial s + (1/s)\partial A_\phi/\partial \phi$ for a vector $\mib{A}$. 
We separate the variables, \mib{u} and \mib{B},
 into the temporal mean (denoted by tildes) and fluctuation (by primes) in time. 
Linearising (\ref{eq:z-mean_vorticity_eq}), together with (\ref{eq:induction}), 
 yields
\begin{equation}
 \frac{D^2 \langle \xi'_z \rangle}{Dt^2}
  + \beta \frac{D \langle u'_s \rangle}{Dt}
 = \frac{1}{\mu_0 \langle\overline{\rho}\rangle} 
    \left\langle ( \widetilde{\mib{B}} \cdot \nabla_\text{H} )
                 ( \widetilde{\mib{B}} \cdot \nabla_\text{H} ) \xi'_z \right\rangle   \;,
   \label{eq:z-mean_vorticity_eq_linear}
\end{equation}
provided ${\partial \langle \overline{\rho} \xi_z \rangle /\partial t}  \sim  \langle \overline{\rho} \rangle {\partial \langle \xi'_z \rangle/\partial t}$.
Now the Coriolis term is represented via the beta parameter.
For the incompressible/Boussinesq fluids this is given by the topographic effect,  
\begin{eqnarray}
 \beta =   - \frac{2\Omega}{H} \frac{dH}{ds} 
  \; , \label{eq:beta_topo0}
\end{eqnarray}
 as non-penetrative conditions at the boundaries imply
 $u_z = \pm u_s\,{dH/ds} = \mp u_s\,{s/H}$ at $z=\pm H$ outside the tangent cylinder \citep[e.g.][]{Bus70}.
When the density varies significantly, \revtwo{the beta effect} instead arises from compressible effects,
\begin{eqnarray}
 \beta = - \frac{2\Omega}{\langle\overline{\rho}\rangle} 
             \frac{d \langle \overline{\rho} \rangle}{d s}  \;, 
  \label{eq:beta_comp} 
\end{eqnarray}
where the $z$-integral of the third term of (\ref{eq:z-mean_vorticity_eq}) is performed \citep[e.g.][]{GHW14,STIY18}. 
Here note the validity of this expression depends on the dynamics of the anelastic fluid,
 particularly on \revtwo{the $z$-integral of the Coriolis term, 
 i.e. to what extent the vorticity is stretched along $z$ between the boundaries $\pm H$.
For more discussions on compressible beta effects we refer to \citet{GER09,JKM09,VS14,BS14}. }
In Earth's fluid core in which the density change is minor and there are solid boundaries, 
 the topographic effect is clearly a reasonable driver. However, 
the compressible effect will be relevant in Jupiter's interior.

\quad 

Introducing the streamfunction $\psi$ for the velocity perturbation, 
 e.g. $\langle \mib{u}' \rangle \sim \nabla_\text{H} \times \psi (s,\phi,t) \hat{\mib{e}}_z$
 with $\hat{\mib{e}}_z$ being the unit vector in the direction of the rotation axis, 
 we find (\ref{eq:z-mean_vorticity_eq_linear}) to give a wave equation. 
We now suppose that the background magnetic field and flow are both \revtwo{steady and axisymmetric
 to seek solutions of the form of $\psi = \hat{\psi} \exp{\mathrm{i} (m\phi - \omega t)}$.
Here the background flow is supposed to be dominated by the zonal component.} 
The wave equation then becomes  
\begin{equation}
\small
 \frac{1}{\mu_0 \langle \overline{\rho} \rangle} 
   \left\langle 
    \overline{\widetilde{B_s}} \frac{d}{ds}
    \overline{\widetilde{B_s}} \frac{d}{ds} 
     \left( \frac{1}{s}\frac{d}{ds} s\frac{d}{ds}
     - \frac{m^2}{s^2} \right) \hat{\psi}
   \right\rangle
 + (\hat{\omega}^2 - \omega_\text{M}^2 )
   \left(  
      \frac{1}{s}\frac{d}{ds} s\frac{d}{ds} 
    - \frac{m^2}{s^2}
   \right) \hat{\psi}
 + \frac{\beta \hat{\omega} m}{s}  \hat{\psi} = 0 ,
    \label{eq:z-mean_vorticity_eq_linear_axisym}
\end{equation}
where 
$\hat{\omega} = \omega - {\langle \overline{\widetilde{U_\phi}} \rangle m/s }$, and 
 \revone{the squared Alfv\'{e}n frequency is given by} $\omega_\text{M}^2 = { \langle \overline{\widetilde{B_\phi^2}} \rangle m^2/\revone{\mu_0} \langle \overline{\rho} \rangle s^2 }$.
This equation goes singular
 when $\overline{\widetilde{B_s^2}}/\mu_0 \langle \overline{\rho} \rangle$,
 the local speed of torsional Alfv\'{e}n waves, crosses zero. 
If the torsional wave is slow compared with the Alfv\'{e}n and Rossby waves travelling in azimuth, equation~
 (\ref{eq:z-mean_vorticity_eq_linear_axisym}) may be further reduced to a \revtwo{second-order} ODE: 
\begin{equation}
 (\hat{\omega}^2 - \omega_\text{M}^2 )
  \left( 
    \frac{1}{s}\frac{d}{ds} s\frac{d}{ds} 
   -\frac{m^2}{s^2}
  \right) \hat{\psi}
 +  \frac{\beta \hat{\omega} m}{s}  \hat{\psi} = 0 \; .
  \label{eq:ode_axis-azimuth-field}
\end{equation}
Here a critical layer will appear if $\hat{\omega}^2 \rightarrow \omega_\text{M}^2$.
If this does not occur in the domain, 
 (\ref{eq:ode_axis-azimuth-field}) yields a set of two solutions. 
This eigenvalue problem for different profiles of $\omega_\text{M}^2$ in Boussinesq fluids
 was explored by \citet{CFF14}.

\begin{figure}[h]%
\centering
 \begin{tabular}{ll}
  (a) & \hspace{2mm}(b) \\
  \includegraphics[viewport=-5mm 1mm 180mm 91mm,clip,width=0.5\linewidth]{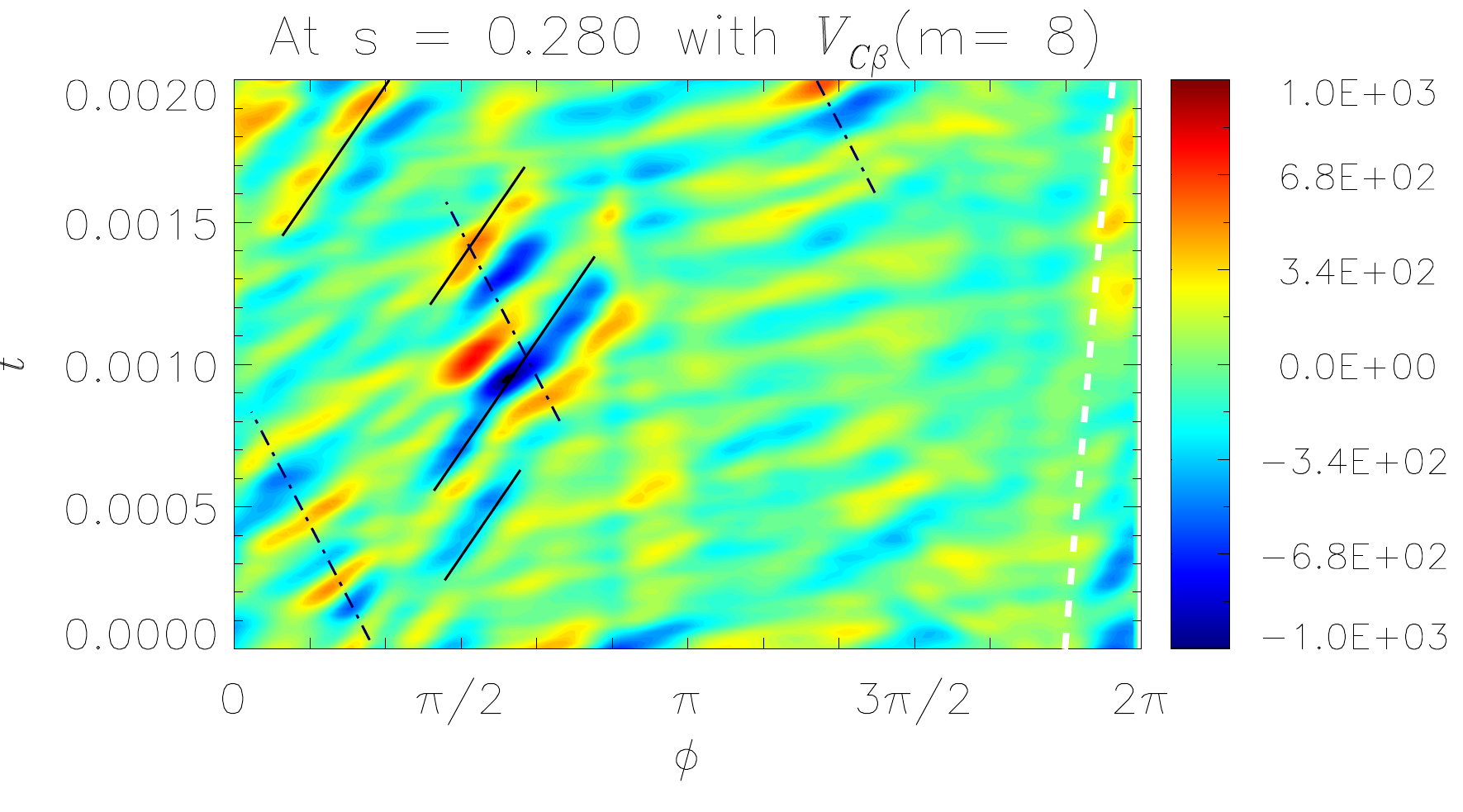} &
  \includegraphics[viewport=0mm 1mm 180mm 91mm,clip,width=0.5\linewidth]{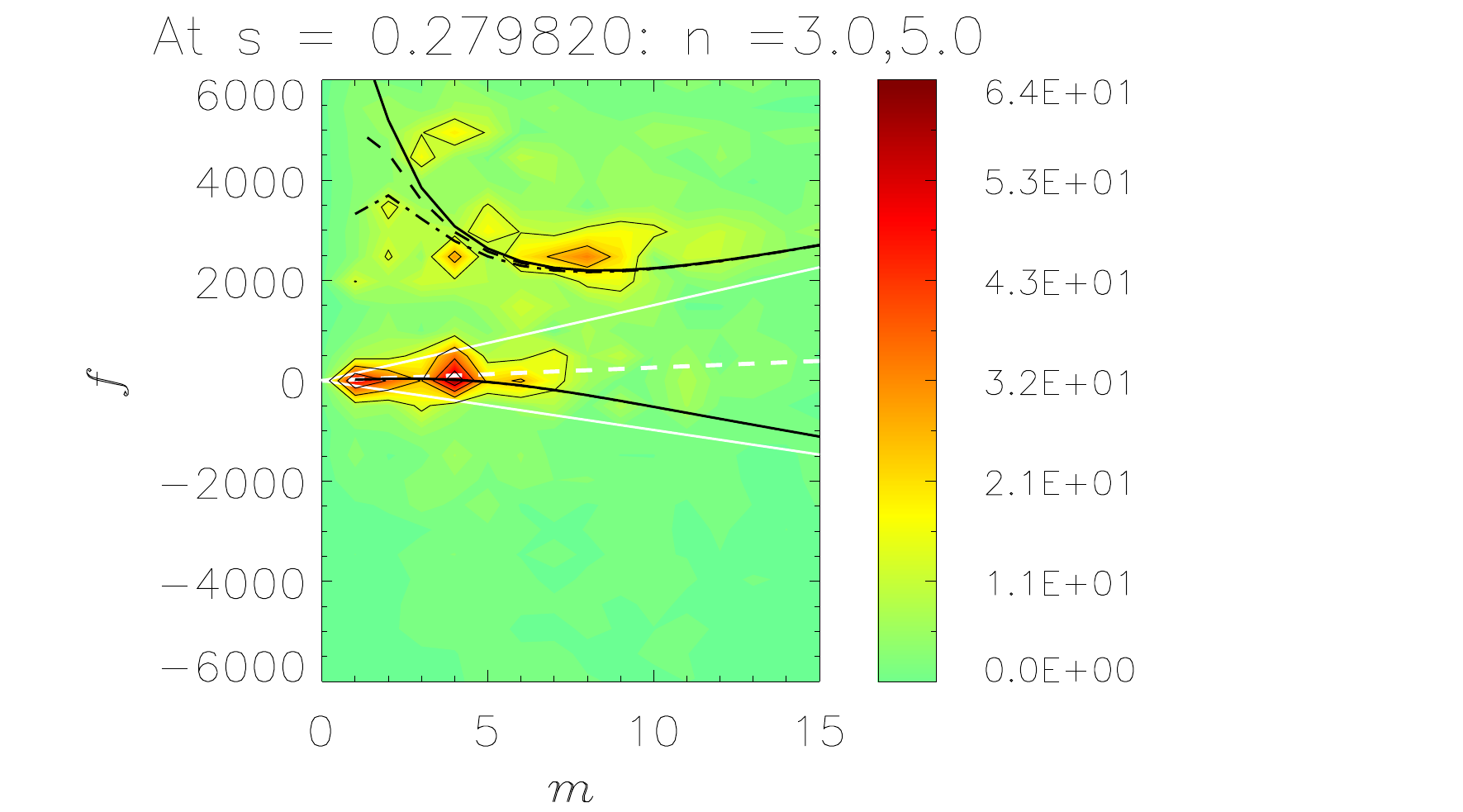} \\
% \\
  (c) & \hspace{2mm}(d) \\
  \includegraphics[width=0.47\linewidth]{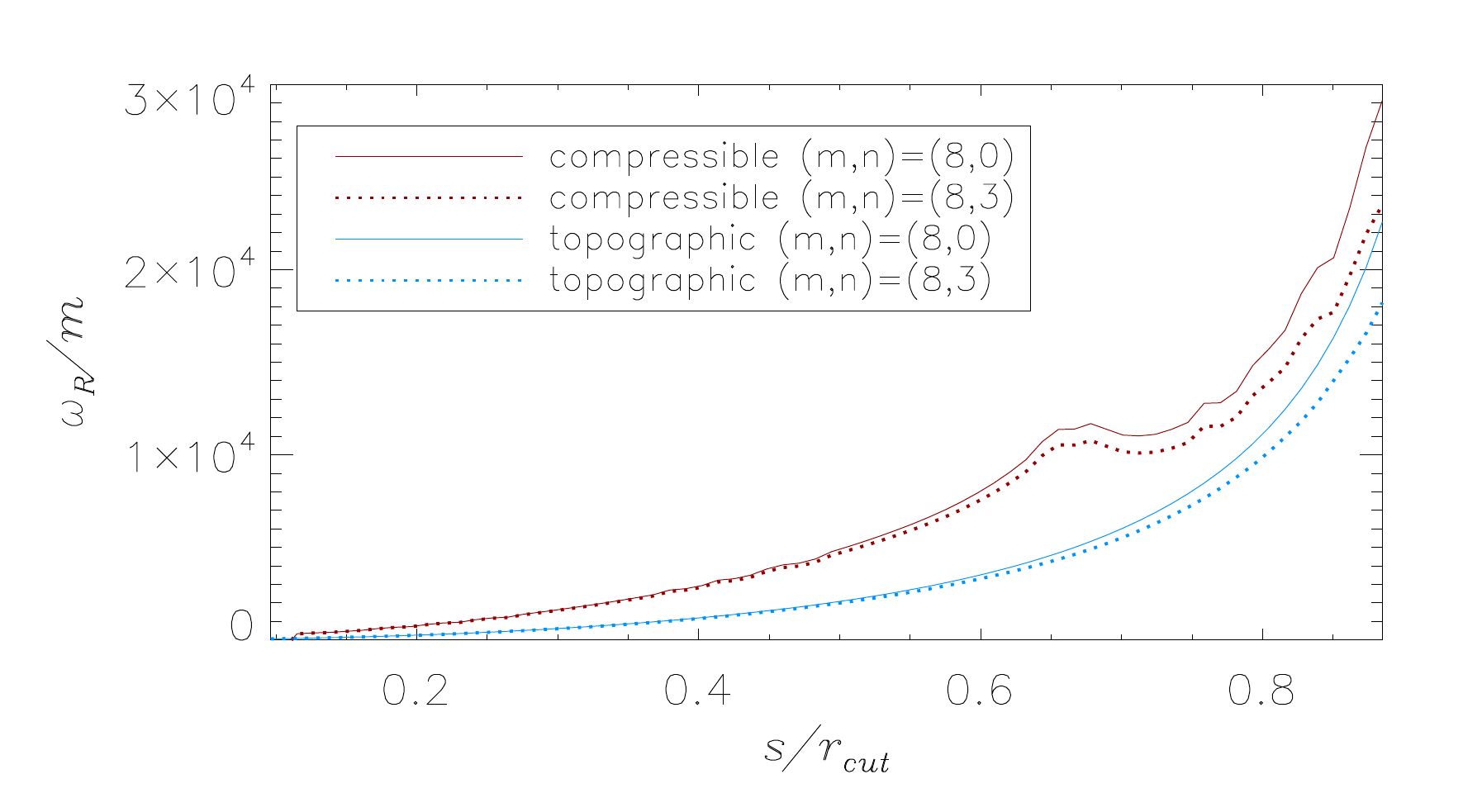} &
  \hspace{2mm}\includegraphics[width=0.42\linewidth]{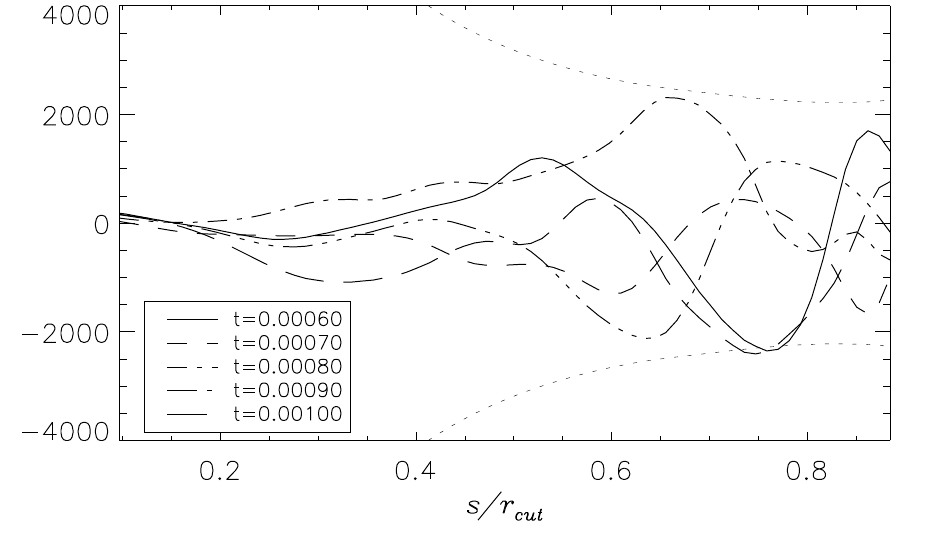} 
 \end{tabular}
 \caption{Fast magnetic Rossby waves seen in a jovian dynamo simulation (run E in \citet{J14,HTJ19}).
 (a) Azimuth time section and (b) wavenumber-frequency power spectrum of $\langle u'_s \rangle$
 at $s = 0.25 r_\text{cut} \sim 0.24 R_\text{J}$.
 In figure a, solid black lines represent phase paths of fast compressible Rossby waves plus the zonal flow advection, $\omega_{+}/m + \langle \overline{\widetilde{U_\phi}} \rangle/s$, for $m = 8$ and $n=0$:
 dashed-dotted lines indicate their group velocity, $\partial\omega_{+}/\partial m + \langle \overline{\widetilde{U_\phi}} \rangle /s$.
 In figure b, black curves show the expected dispersion relations of advection plus wave $\omega_\pm$ for the compressible beta parameter and $n=0$ (solid), $3$ (dashed), and $5$ (dashed-dotted). 
 White dashed lines for the advection only; 
 white solid curves for the advection plus the Alfv\'{e}n wave ($\pm \omega_\text{M}$).
 (c) Phase speeds of compressible (red) and topographic (blue) Rossby waves,
 $\omega_\text{R}/m$, as a function of normalised $s$.
 (d) Radial profiles of $\langle u'_s \rangle$ at $\phi \sim 2\pi/3$.
 Curves exhibit snapshots at different times, which are indicated in the legend. 
 Dotted ones indicate the expected variability $s^{-3/2}$ \revone{[see Appendix~\ref{sec:WKBJ} for details]}.\\}
 \label{fig:MRW_DNS_jup}
\end{figure}

To examine the basic properties of the equation, we here suppose a WKBJ-type solution, 
 \revtwo{$\hat{\psi} = A_0 s^{-1/2} \exp{ \mathrm{i} \int n(s)ds}$}
 [see Appendix~\ref{sec:WKBJ} for details],
where the local dispersion relation is given by 
\begin{equation}
 \hat{\omega}^2 - \hat{\omega} \; \omega_\text{R} - \omega_\text{M}^2 = 0 \; , \label{eq:dispersion_relation}
\end{equation}
and the Rossby wave frequency 
$\omega_\text{R} = \beta m s/(m^2 + n^2 s^2  + 1/2)$.
The quadratic equation (\ref{eq:dispersion_relation})
 has roots 
\begin{equation}
 \omega_\pm = \omega_\text{R} \left[ \frac{1}{2}
	     \pm \frac{1}{2} \sqrt{1 + 4\frac{\omega_\text{M}^2}{\omega_\text{R}^2}} \right] \; .
\end{equation}
In the limit $\omega_\text{M}^2/\omega_\text{R}^2 \gg 1$, e.g. high wavenumbers for a given basic state,  
 these simply yield the Alfv\'{e}n waves along the toroidal field. 
Their unique properties become evident in another limit $\omega_\text{M}^2/\omega_\text{R}^2 \ll 1$
 to yield
 \begin{equation}
  \omega_{+}
   \sim
      \omega_\text{R} \left( 1 + \frac{\omega_\text{M}^2}{\omega_\text{R}^2} \right)
  \; \quad \textrm{and} \quad  
  \omega_{-}
   \sim -\frac{\omega_\text{M}^2}{\omega_\text{R}}  \; . \label{eq:fast}
 \end{equation}
The fast modes, with frequency $\omega_{+}$, are essentially equivalent to the hydrodynamic waves.
Their timescales are basically ruled by $\beta$, or the planet's rotation rate, 
 but are shorter in the presence of the background magnetic field.
Their phase velocity is prograde in a thick shell problem (such as that applicable for the Earth's fluid core) \citep{Bus70,Bus86},
 while the group velocity is retrograde.
Note that these directions appear to be opposite from the conventional Rossby waves in the atmosphere. 
Figure~\ref{fig:MRW_DNS_jup} demonstrates a fast wave seen in Jovian dynamo simulations,
 in which the compressible beta effect plays a role.

The slow modes, with frequency $\omega_{-}$, are unique to the rotating MHD system,
 travelling retrogradely; the frequency is
 given by the ratio of the squared Alfv\'{e}n frequency to the \revtwo{Rossby frequency}. 
Hence they are sensitive to $\langle \overline{\widetilde{B_\phi^2}} \rangle$, or the toroidal field strength. 
Their timescales may vary from $10^1$ to $10^4$ years in Earth's fluid core,
 indicating a link to the centennial geomagnetic westward drift \citep{H66} (sec.~\ref{sec:MRW_geo}). 
The dispersive nature of this mode is also noticeable. 
For the simpler case of constant density, 
 the slow mode dispersion relation (\ref{eq:fast}) may be rewritten as 
 \begin{equation}
  \omega_{-} \sim
     - \frac{\langle \overline{\widetilde{B_\phi^2}} \rangle (r_\text{o}^2 - s^2)}
            {2 \mu_0 \langle \overline{\rho} \rangle \Omega s^4}
       m (m^2 + s^2 n^2) \; ,\label{eq:slow_approx}
 \end{equation}
 where the geometrical effect \revtwo{in the Rossby frequency} is omitted. 
This is reduced to a relationship proportional to $m^3$
 when the azimuthal wavenumber dominates over the radial one.
The wave motion is highly dispersive in the $\phi$ direction; this dispersive nature gives a strong steer to possible nonlinear behaviour (i.e. the presence of solitons as discussed in sec.~\ref{sec:MRW_nonlinear}).
This is not the case when the radial structure is more complicated\revone{, i.e. $m^2 \ll s^2 n^2$};
in that case
 the $\phi$-propagation is largely non-dispersive
 while the $s$-propagation is weakly dispersive. 
\revtwo{We here recall that both magnetic Rossby modes are capable of travelling in $s$: this is analogous to the atmospheric version which may travel in latitude too \citep[e.g.][]{V17}.}

%%%% new Figure 8 %%%%%%%%%%%%%%%%%%%%%%%%%%%%%%%%%%%%%%%%%%%%
\bigskip
%%%%%%%%%%%%%%%%%%%%%%%%%%%%%%%%%%%%%%%%%%%%%%%%

%\quad

%
All theory needs to be re-addressed 
 when ${\left\|\overline{\widetilde{B_s}} \partial /\partial s \right\| \gg \left\| (\overline{\widetilde{B_\phi}}/s) \partial/\partial \phi \right\|}$.
As indicated from (\ref{eq:z-mean_vorticity_eq_linear}) or (\ref{eq:z-mean_vorticity_eq_linear_axisym})
 the slow mode for the case would imply highly dispersive motion in $s$.
This seems to be the regime recently explored by \citet{GJN21},
 who computed eigenmodes in an extended 2d model for a nonaxisymmetric background $\widetilde{B_s}$
 to obtain high wavenumber modes for the interannual westward drift (sec.~\ref{sec:MRW_geo}).

\quad

\subsection{Magnetic Rossby waves in the Earth}  \label{sec:MRW_geo}

Slow magnetic Rossby waves were proposed by \citet{H66}
 to explain the $\sim$300 year geomagnetic westward drift (sec.~\ref{sec:intro}). 
This migration has been seen in centennial models \citep[e.g.][]{FJ03}
 and in millennial models \citep[e.g.][]{HG18,NSKHH20}. 
A complementary scenario to this is that the westward drift arises because of the advection by large-scale flows in the geodynamo
 \citep[e.g.][]{BFGN50,AFF13}.
It is more likely that the observed feature consists of a combination of advection and wave propagation;
some early works on numerical dynamos pointed out that
 migration speeds seen in simulations did not match the flow advection speed \citep{KR02,CO03}.

\begin{figure}[t]%
\centering
 \begin{tabular}{ll}
  (a) & \hspace{5mm}(b) \\
  \includegraphics[viewport=0mm 1mm 285mm 141mm,clip,width=0.5\linewidth]{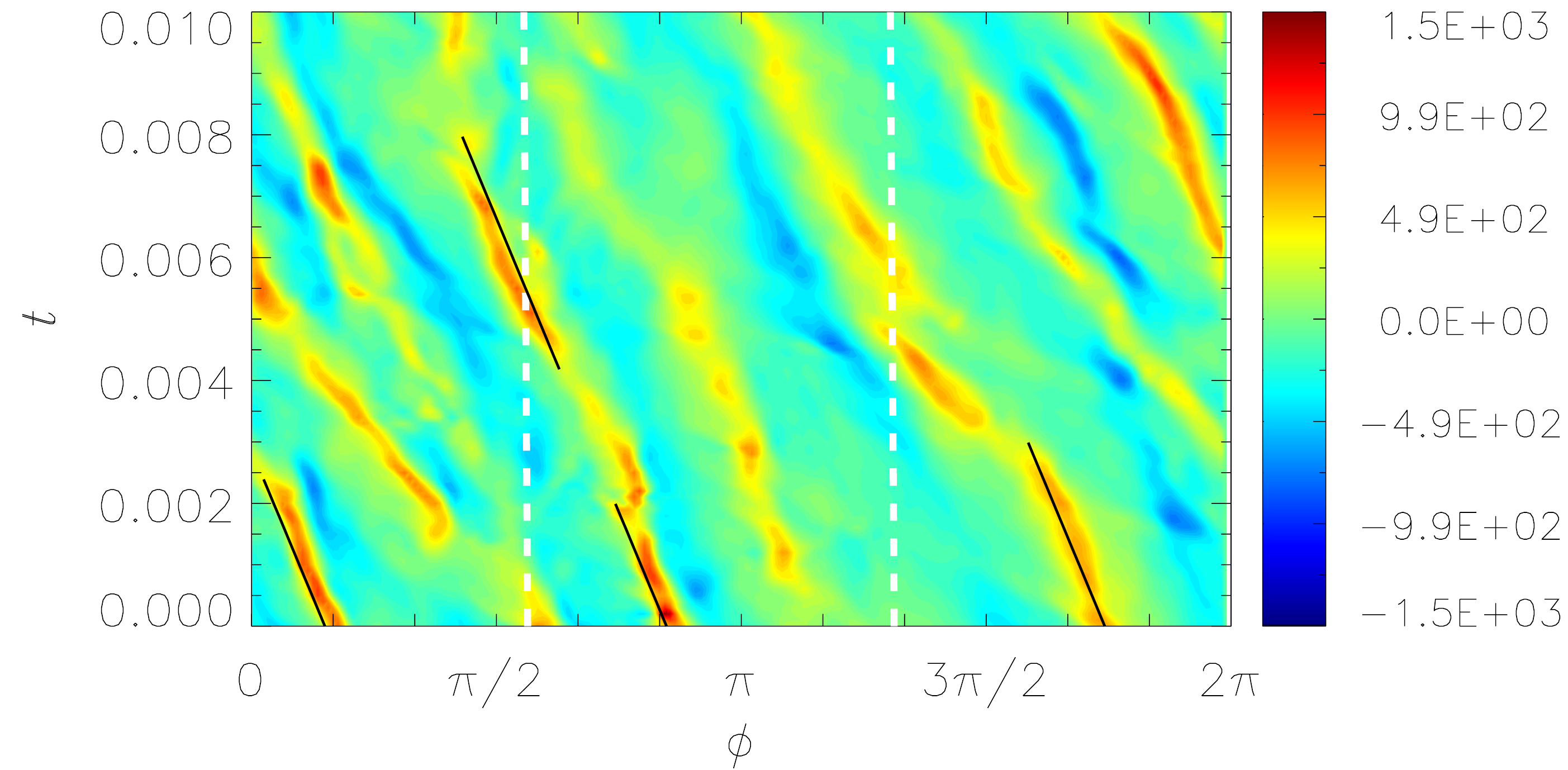} &
  \includegraphics[width=0.45\linewidth]{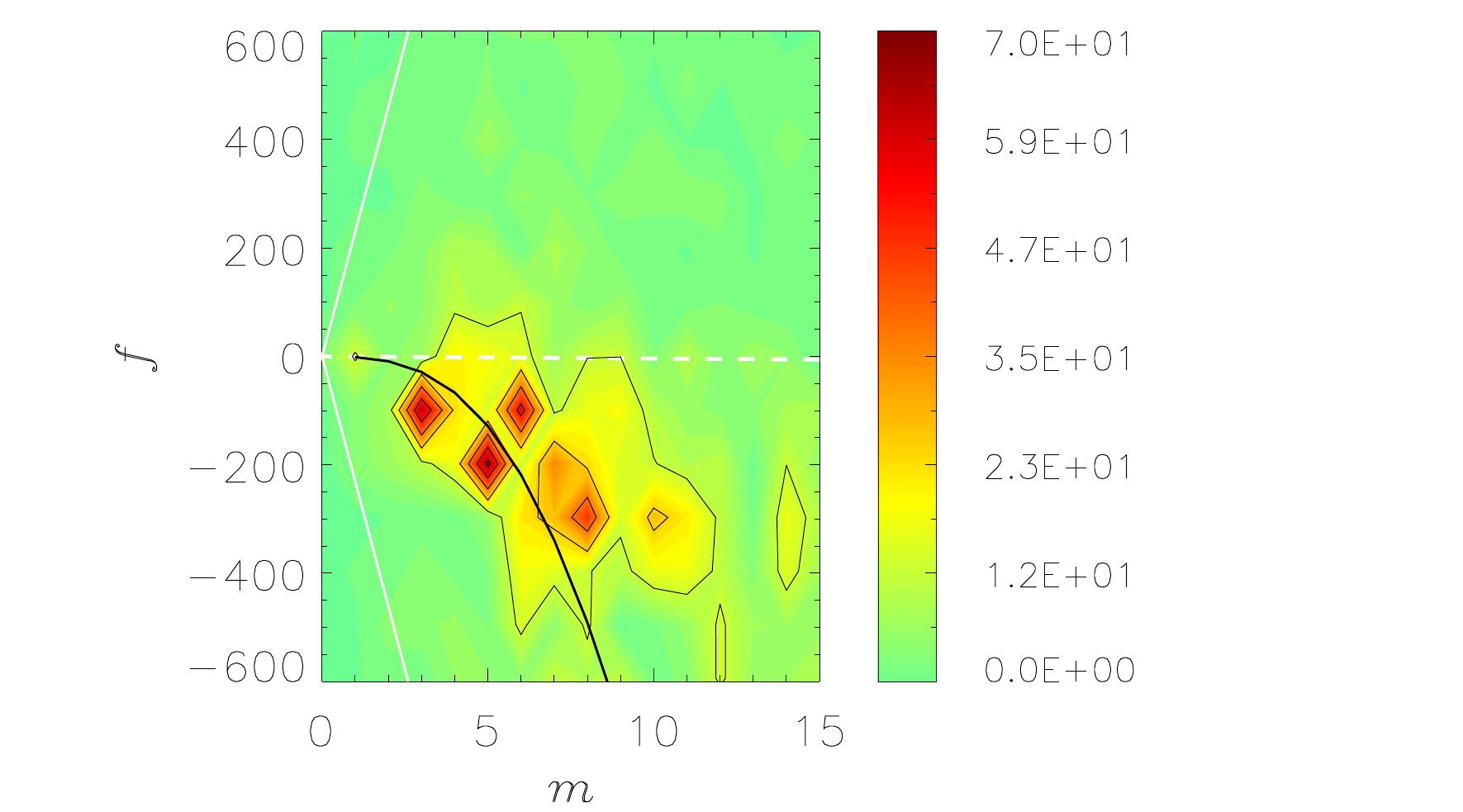}
 \end{tabular}
 \caption{Slow magnetic Rossby waves in a geodynamo simulation
 (adapted from \citet{HTJ18}).
 (a) Azimuth time section and (b) wavenumber-frequency power spectrum of $\langle u'_s \rangle$
 at the mid radius $s = 0.5 r_\text{core}$.
 In figure a solid black lines represent phase paths of slow magnetic Rossby waves
 plus the zonal flow advection, i.e. $\omega_{-}/m + \langle \overline{\widetilde{U_\phi}} \rangle /s$ for $m = 5$ and $n=0$; white dashed lines indicate the advection speed only. 
 In figure b, solid black (white) curves represent the expected dispersion relations of advection plus the wave $\omega_\pm$ ($\omega_\text{M}$) for the topographic beta parameter  $n=0$. White dashed lines for the advection only.\\}
 \label{fig:MRW_DNS}
\end{figure}

Using updated geodynamo simulations,
 \citet{HJT15} re-addressed nonaxisymmetric motions in terms of the 2d theory above, 
 and demonstrated that the retrograde drifts \revtwo{in the simulations} were well explained by slow waves
 (\ref{eq:slow_approx}) riding on the mean flow advection
 $\langle \overline{\widetilde{U_\phi}} \rangle m/s$
 (figure~\ref{fig:MRW_DNS}). 
The nonaxisymmetric waves may be excited through any driving mechanism; 
 here convection in the spherical shell plays a major role. 
The preferred wavenumber, or the frequency, is thus determined by convective activities. 
The nature of the observed waves clearly depends on the regime of the driving mechanism:
 in the case of convection a slow wave will be favourable
 when the magnetic diffusion time is longer than the thermal one, 
 while the opposite regime will yield other modes including diffusive modes travelling progradely \citep{Bus76,F08,HTS14}.
Analyses of the simulations confirmed that 
 the slow waves emerged when the magnetostrophic terms were dominant in the
 vorticity equation (\ref{eq:z-mean_vorticity_eq}) \citep{HTJ18}.
The identification of those waves, as well as torsional Alfv\'{e}n waves,
 may signify a dynamo in the magnetostrophic regime.

%%%% new Figure 9 %%%%%%%%%%%%%%%%%%%%%%%%%%%%%%%%%%%%%%%%%%%%
\bigskip
%%%%%%%%%%%%%%%%%%%%%%%%%%%%%%%%%%%%%%%%%%%%%%%%

%\quad 

\begin{figure}[t]%
\centering
 \begin{tabular}{ll}
 (a) \hspace{0.5\linewidth} (b)& \\
 \hspace{2mm}\includegraphics[width=0.95\linewidth]{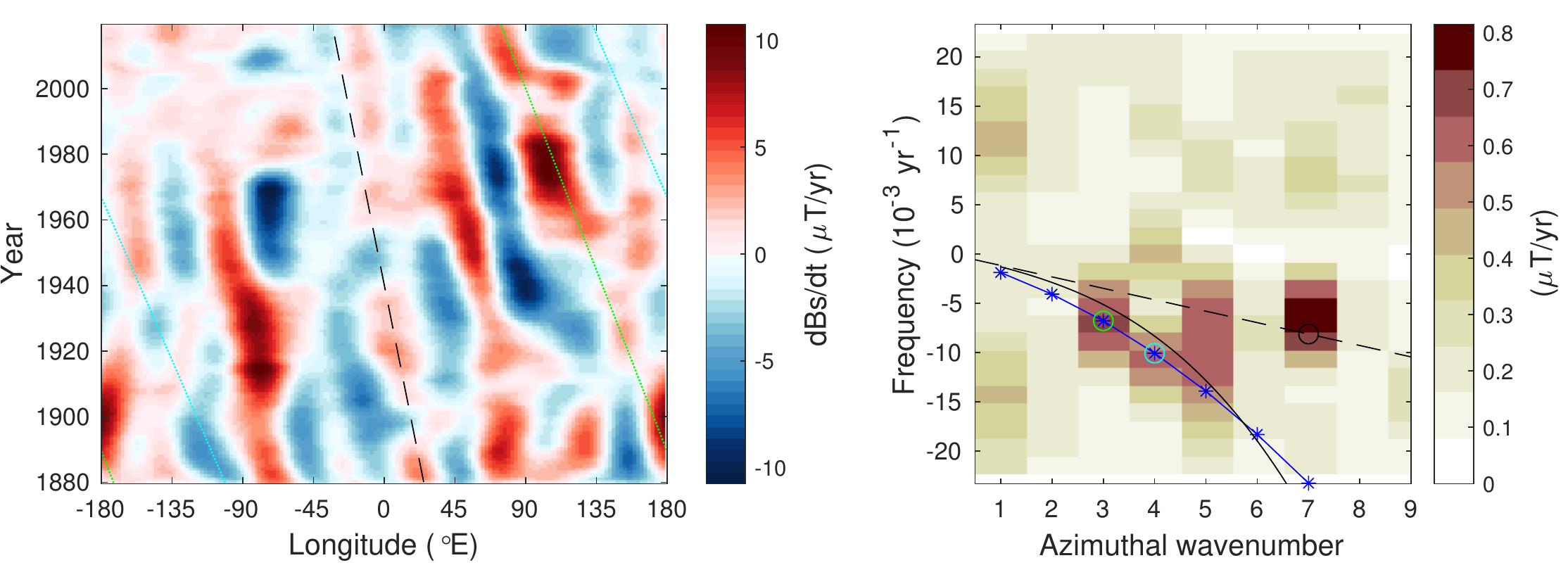} 
 \end{tabular}
  \caption{Nonaxisymmetric motion of geomagnetic secular variation,
  $\partial B_s/\partial t$, in 1880-2015 %(cov-obs2019 \citep{GHA19}).
  (\revone{produced from \citet{GHA19}}). 
 (a) Longitude-time section, H\"{o}vmoller diagrams, and (b) 2d spectrum, the sum over 39-41$^\circ$N. 
 In figure a the green, cyan, black lines indicate the speeds of signals at $m = 3$, $4$, and $7$,
 respectively, as identified by circles in the spectrum. 
 In figure b the dashed line represents $f_\text{adv} = U_0 m/2\pi s$
 given $U_0 = 0.32^\circ$/yr at the radius $s=0.77 r_\text{core}$ \citep{PMS15,HJT15}; 
 the black solid curve represents $f = f_\text{adv} + \omega_{-}/2\pi$, based on the local theory
 (\ref{eq:slow_approx}) for ${\langle \overline{\widetilde{B_\phi^2}} \rangle^{1/2} } =$ 15\,mT; 
 blue asterisks indicate $f = f_\text{adv}$ plus the frequency of the first normal mode solutions
 (\ref{eq:ode_axis-azimuth-field})
 for Malkus field ${\langle \overline{\widetilde{B_\phi}} \rangle } \propto s$ of magnitude 13mT
 (fig.~\ref{fig:TW_normal_modes}b).\\} 
 \label{fig:MRW_geomag}
\end{figure}

It is useful to examine the geomagnetic data for nonaxisymmetric components.
Figure~\ref{fig:MRW_geomag}a displays the longitude-time section, from 1880 to 2015, of the secular variation,
 $\partial B_s/\partial t$, at latitude $\sim$40$^\circ$N corresponding to $s/r_\text{core} \sim 0.77$,
 in cov-obs2019. 
The westward drift, clearly visible on this timescale, appears to consist of multiple drift speeds. 
This is evident by the 2d FFT spectrum in figure b. 
Here a linear relation, $\omega \propto m$, indicated by the dashed line,
 represents an advection effect by mean flow. 
Clearly this simple advection model can not  explain the multiple signals observed.  
We add the dispersion relations of the slow wave too: 
 the black solid curve for the local theory (\ref{eq:slow_approx})
 for ${ \langle \overline{\widetilde{B_\phi}} \rangle^{1/2} } \sim$ 15 mT
 and blue asterisks for normal mode solutions
 provided a background ${\overline{\widetilde{B_\phi}} } \propto s$ of maximum 13 mT
 (Appendix~\ref{sec:normal_modes} and fig.~\ref{fig:TW_normal_modes}b).
Those speeds for chosen $m$ are indicated by different lines in the figure a. 
This attempt is inconclusive but indicative
 that today's geomagnetic datasets are capable of capturing the signatures of waves.
It would be crucial to analyse them on multiple timescales;
the slow wave timescale may vary by a few orders of magnitude (see above).
Probing the slow wave will enable the estimation of the toroidal magnetic field  \citep{HJT15},
 which is confined within the dynamo region\revtwo{, i.e. inaccessible through direct measurements}.

\quad 

Beyond the framework above,
 a zoo of nonaxisymmetric waves is being explored. 
State-of-the-art numerical geodynamo calculations
 exhibit different wave classes \citep[e.g.][]{AF19,AG21}
 such as Alfv\'{e}n modes about inhomogeneous poloidal part $\widetilde{B_s}$
 and also fast Rossby modes likely. 
These have been linked
 geomagnetic jerks (sec.~\ref{sec:intro})
 and to nonlinear interactions with the convective dynamics for the dynamo. 
More recently, based on the linear calculations by \citet{GJN21},
 \citet{GGJetal22} attributed slow modes of high radial wavenumber
 for a background $\widetilde{B_s}$ 
 to the equatorial westward drift of $\sim$6 years (sec.~\ref{subsec:obs}). 
An alternative idea for the rapid drift is that
 Rossby waves are excited in the stratified layer at the top of the core, as an MAC wave \citep{BM19}.
These are ongoing topics; we shall remark further in the final section.

%%%% new Figure 10 %%%%%%%%%%%%%%%%%%%%%%%%%%%%%%%%%%%%%%%%%%%%
\bigskip
%%%%%%%%%%%%%%%%%%%%%%%%%%%%%%%%%%%%%%%%%%%%%%%%

%\quad

\subsection{Finite amplitude effects}  \label{sec:MRW_nonlinear}

A novel finding by the numerical simulations described above was the sharp waveform of the slow wave (e.g. figure~\ref{fig:MRW_DNS}a). 
These are isolated and steepened,
 rather than forming wave trains as expected for a linear dispersive wave.
Moreover, their crests appear to be cleaner than the troughs. 
Those observations are reminiscent of cnoidal waves and solitons of finite amplitude,
 which are both known to be solutions of the Korteweg-de Vries (KdV) equation.
Indeed, the approximated dispersion relation (\ref{eq:slow_approx}) has the dispersive term 
 proportional to $m^3$, as in the KdV equation.
Weakly nonlinear analyses were recently explored \citep{H19,HTJ20}
 in terms of 2d annulus models \citep{Bus76} and spherical models \citep{CFF14}.
This contrasts with analyses in equatorial shallow-water MHD \citep{L17},
 in which fast modes in a stratified environment were a primary focus. 
(Hydrodynamic Rossby waves are known to shape coherent structures
 and to be governed by soliton equations in certain regimes \citep[e.g.][]{R77,WY84};  
 those solutions were proposed as an explanation for the Jupiter's Great Red Spot.)

\quad

The model setting adopted by \citep{H19,HTJ20} is essentially same as above (figure~\ref{fig:MRW}a).
For simplicity the magnetic field is also assumed to be two dimensional
 so that it can be represented
 by the magnetic potential $g$ such that $\mib{B} = \nabla \times g (s,\phi,t) \hat{\mib{e}}_z$; 
 this is analogous to the streamfunction $\psi$ for the velocity,
 $\mib{u} = \nabla \times \psi (s,\phi,t) \hat{\mib{e}}_z$.
Also we suppose the density is constant and the beta parameter is topographic. 
Following a standard multiple-scale technique called the reductive perturbation method, 
 we introduce slow variables with small perturbation $\epsilon$ ($\ll 1$) such that 
 $\tau = \epsilon^{3/2} t$ and $\zeta = \epsilon^{1/2}(\phi - ct)$
 and expand the two variables
 to get asymptotic solutions such that $[\psi , g] = [\psi_0, g_0] + \epsilon [\psi_1, g_1] + ..$.
Hence a long-wave limit is being studied.

The zeroth order is given by the basic state. 
At the first order $\mathcal{O}(\epsilon)$ the two governing equations
 yield a linear, 2nd-order homogeneous PDE for $g_1$.
Assuming a separable solution in form $g_1 = \Phi (s) G(\zeta,\tau)$
 reduces the problem to an ODE in dimensionless form, 
\begin{equation}
 \mathcal{L} \Phi \equiv 
   \left\{ 
            \frac{\overline{\widetilde{B_\phi}}}{ \beta s} \left[ 
		    \frac{ \overline{\widetilde{B_\phi}} }{s} \frac{d}{ds} s \frac{d}{ds}
		    - \frac{d}{ds} \frac{1}{s} \frac{d}{ds} s \overline{\widetilde{B_\phi}}
							   \right]  
   +  \left( \frac{\overline{\widetilde{U_\phi}} }{s} - c \right) 
   \right\} \Phi  = 0   \;,
   \label{eq:g1_pde_sphere}
\end{equation}
where $\mathcal{L}$ denotes the linear differential operator
 comprising of $s$, $d/ds$, $\overline{\widetilde{B_\phi}}$, 
 $\beta, \overline{\widetilde{U_\phi}}$, and $c$. 
This is an eigenvalue problem with eigenvalues $c$ and associated eigenfunctions $\Phi$, 
 together with appropriate boundary conditions.
Here it is worth noting that the equation becomes singular as $\overline{\widetilde{B_\phi}}^2/\beta \rightarrow 0$
 but this is unlikely as $\overline{\widetilde{U_\phi}}/s \rightarrow c$. 
This is distinct from the hydrodynamic cases; there
 \citet{R77} addressed solitary Rossby waves in the vicinity of the critical layer
 when a wave speed approaches the mean flow speed. 
What happens around any magnetic critical layer, including its continuous solutions, 
 is entirely uncertain.

Focusing on the discontinuous solutions,
 we proceed to the next order to determine the structural function $G (\zeta, \tau)$.
After some algebra, the vorticity and induction equations at $\mathcal{O}(\epsilon^2)$
 are found to yield an inhomogeneous PDE for $g_2$,
 whose homogeneous part is given as $\mathcal{L}g_2 = 0$. 
We thus require a solvability condition to suppress the secular terms, yielding 
\begin{equation}
 \frac{\partial G}{\partial \tau}
 + \alpha \; G \frac{\partial G}{\partial \zeta}
 + \gamma \; \frac{\partial^3 G}{\partial \zeta^3} = 0 \; . 
   \label{eq:KdV}
\end{equation}
Here $\alpha$ and $\gamma$ are determined from the $\mathcal{O}(\epsilon)$-eigenfunction $\Phi$,
 its adjoint solution $\Phi^\dag$, and the basic state $\overline{\widetilde{B_\phi}}$,
 $\overline{\widetilde{U_\phi}}$, and $\beta$:
 see detailed expressions in \citet{HTJ20}.
This evolution of the structural function $G (\zeta, \tau)$, and hence $g_1$,
 is therefore governed by the the Korteweg-de Vries equation if the coefficients are both nonzero. 
Equivalent analyses in the cartesian model \citep{H19} show that
 the coefficient of nonlinear effect would be nonzero unless $\overline{\widetilde{B_\phi}}$,
 $\beta$, and $\overline{\widetilde{U_\phi}}$ are all uniform. 
This could be readily satisfied for a spherical system, for which $\beta$ is nonuniform.

\begin{figure}[t]%
\centering
\begin{tabular}{lll}
 \hspace{-4mm} (a) &  \hspace{10mm}(b) \\
 \includegraphics[width=0.25\linewidth]{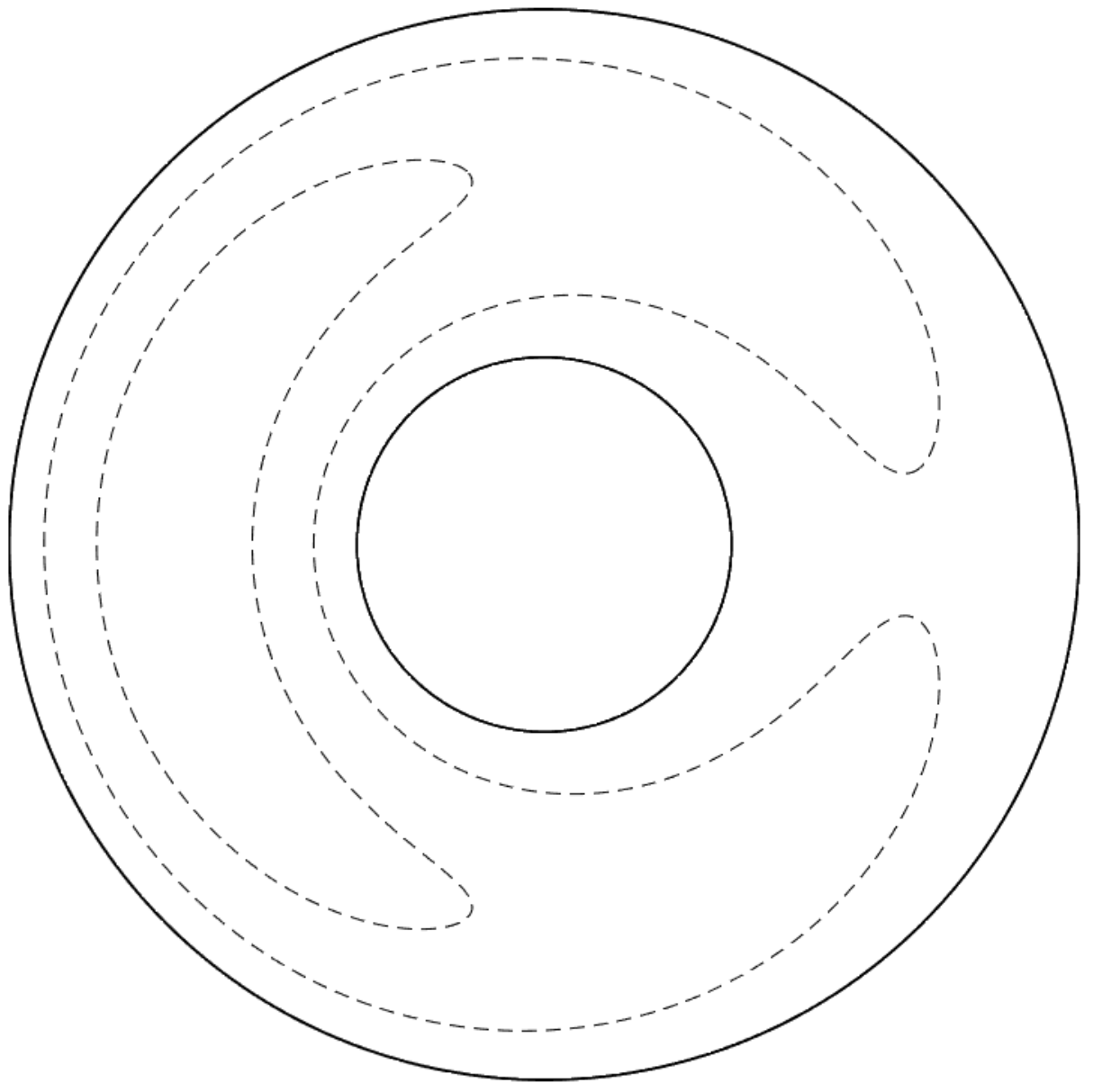} &
 \hspace{10mm}
 \includegraphics[width=0.25\linewidth]{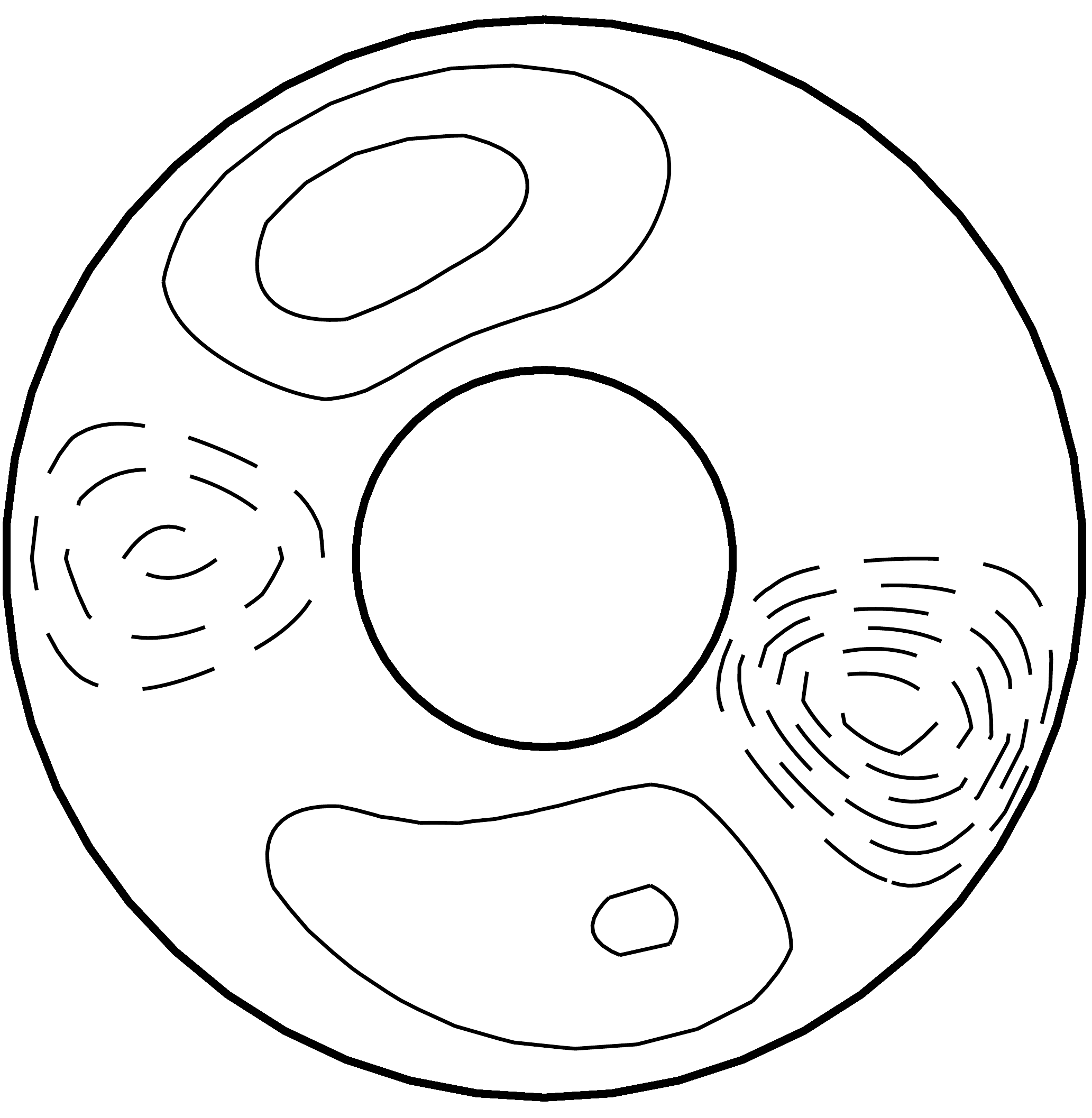} 
 \end{tabular}
 \caption{(a) 1-soliton and (b) N-soliton solutions of slow magnetic Rossby waves for Malkus field,
 $\overline{\widetilde{B_\phi}} = s$ (after \citet{HTJ20}).
 Streamfunction $\psi$ in the equatorial plane in snapshots. Dashed (solid) curves indicate their negative (positive) values,
 implying anticyclonic (cyclonic) motion.\\}
 \label{fig:MRW_nonlinear}
\end{figure}

In spherical shells 
 \citet{HTJ20} solved the eigenvalues problem (\ref{eq:g1_pde_sphere}) to calculate the coefficients of (\ref{eq:KdV}) 
 for different sets of the basic state magnetic fields $\overline{\widetilde{B_\phi}}$ and velocity profiles $\overline{\widetilde{U_\phi}}$. 
They found nonzero values for the coefficients for the all cases they explored, implying that the  KdV equation is the correct canonical description. 
As its solutions are well known,  
 our asymptotic solution may be simply illustrated. 
Cases for the 1- and N-soliton solutions are demonstrated in figure~\ref{fig:MRW_nonlinear}.
The solitary wave solution as seen in figure~\ref{fig:MRW_nonlinear}a implies an anticyclonic isolated vortex
 that is drifting retrogradely with the speed of the linear wave, the order of $10^2$ to $10^4$ years in Earth's core.
Here recall that
 core flow inversions have revealed an anticyclonic gyre persisting in the fluid core for more than 100 yrs (fig.~\ref{fig:gyre}). 
An up-to-date geomagnetic model for the past 9000 years was recently reported to exhibit a westward-drifting eastern-western hemispherical asymmetry, with quasi-periodic behaviours of $\sim$1300 years, potentially related to a similar planetary gyre \citep{Netal22}.
The origin of the asymmetry has been discussed
 in terms of couplings with the rocky mantle and the solid inner core \citep[e.g.][]{AFF13}. 
Meanwhile, geodynamo simulations demonstrated the emergence of such a coherent structure as a natural consequence of the fluid dynamics therein \citep{SJNF17}.
The soliton solutions above show that the gyre shape can simply be explained using natural nonlinear wave dynamics.

%%%% new Figure 11 %%%%%%%%%%%%%%%%%%%%%%%%%%%%%%%%%%%%%%%%%%%%
\bigskip
%%%%%%%%%%%%%%%%%%%%%%%%%%%%%%%%%%%%%%%%%%%%%%%%

%\quad

\section{Concluding remarks and perspectives}

In this paper we have discussed the topic of rotating MHD waves, or MC waves,
 largely motivated by recent advances of geophysical observation and numerical modelling.
The subject embraces rich physics in addition to providing the  potential for probing the interiors of natural dynamos.
As illustrated by linear theory, there are many different wave classes. 

\revone{
To obtain fundamental insights, 
 we have paid particular attention to torsional Alfv\'{e}n waves/oscillations and magnetic Rossby waves
 that can be excited in the geo- and jovian- dynamos. 
The two wave classes may be considered as typical modes
 occurring in magnetostrophic balance of rapidly rotating MHD fluids \revtwo{(sec.~\ref{subsec:wave}).
They may particularly be relevant for understanding the planetary magnetic variations, length-of-day variations, and possibly the surface appearance in gaseous planets.
The observations enable the possible inference of the strength and its spatial structure of the poloidal field component within the dynamo region, and of the \revone{``}hidden'' toroidal component there. 
This will provide a crucial constraint on the dynamo theory at all.}
}

%\vspace{5mm}

%
Despite the sterling attempts of theoretical and computational scientists, we believe that there \revtwo{remain} some unexplored issues and open questions, namely: 
\begin{itemize}
 \item The class of rotating MHD waves, whether MC or MAC, is clearly a zoo,
       exhibiting different dynamics and behaviours. 
       Even the slow Rossby wave class appears to be distinct, dependent on the background magnetic field and the regime. 
       We need a concrete catalogue to distinguish these modes and to elucidate their individual behaviours. 
       This will enable the prediction of which waves best suit the inference of the quantity of interest within the dynamo.

 \item A mathematical challenge is
       to address the critical layers arising from the background magnetic field profile,
       the relevance of continuous spectra, and their potential feedbacks on the mean state;
       they were partly addressed \citep{A72,N20}.
       Those concepts have been explored in the plasma physics
       and geophysical fluid dynamics,
       in which mean flows tend to be of primary interest.  
       Their knowledge and techniques could hint at solutions in the current context.

 \item Is it possible to find waves that are topologically protected,
       such as those that have been found to exist in hydrodynamic rotating systems and plasmas \citep{DMV17,PMTZ20,PBMT20}?
       If so, the edge waves could allow us to sense the vicinity of a boundary including a thin stably stratified layer.

 \item From an observational point of view,
       the existence of those waves and their characterisation are still a subject of debate.
       In particular, 
       distinguishing a few candidate modes/branches from data 
       seems to be a tricky issue.
       A methodology to separate individual waves has led to significant progress 
       in Earth's seismology, and meteorology likely.
       Today's data-driven approaches might help to endorse this: 
       they are now capable of extracting signals to incorporate the physics.

 \item Whereas wave motion could provide us the information about deep dynamos, 
       do they play any roles in the dynamo action and the internal dynamics at all?  
       There are classic ideas such as inertial wave generating helicity and thus a dynamo \citep{M78,DR15}
       and the supression of zonal mean flows in the presence of magnetic field \citep{TDH07}.
       Furthermore, the interaction of waves with critical layers could lead to the driving of mean flows. 
       It is uncertain how individual waves classes might feed dynamos. 
       This would be another theoretical challenge for the future.

\end{itemize}

\bigskip

\subsection*{Acknowledgements}
%\bmhead{Acknowledgments}

KH is grateful to C.A.~Jones, R.J.~Teed, Y.-Y.~Hayashi, and A.~Kageyama
 for their insightful comments and supports.
She would also like to thank A.~Sakuraba and C.C.~Finlay for motivating her for the topic. 
Comments by two anonymous reviewers helped to improve the manuscript.
The studies are supported from the Japan Society for the Promotion of Science (JSPS)
 under Grant-in-Aid for Scientific Research (C) No.~20K04104,
 from the Foundation of Kinoshita Memorial Enterprise,
 and from the Swedish Research Council, 2020-04813. 
SMT would like to acknowledge support from the European Research Council (ERC) under the European Unions Horizon 2020 research and innovation program (grant agreement no. D5S-DLV-786780) and by a grant from the Simons Foundation (Grant number 662962, GF)

This work was partly made during the visit at the Isaac Newton Institute for Mathematical Sciences under the programme DYT2 which was supported by EPSRC grant number EP/R014604/1.

%\quad
%\section*{Declarations}

%Some journals require declarations to be submitted in a standardised format. Please check the Instructions for Authors of the journal to which you are submitting to see if you need to complete this section. If yes, your manuscript must contain the following sections under the heading `Declarations':

%\begin{itemize}
%\item Conflict of interest/Competing interests (check journal-specific guidelines for which heading to use)
%\item Ethics approval 
%\item Consent for publication
%\item Availability of data and materials
%\item Authors' contributions
%\end{itemize}

%\noindent
%If any of the sections are not relevant to your manuscript, please include the heading and write `Not applicable' for that section. 

%\subsection*{Conflict of interest}
%%\bmhead{Conflict of interest}
%There is no conflict of interest. 

%\subsection*{Availability of data and materials}
%%\bmhead{Availability of data and materials}
%All data appearing in figs.~\ref{fig:mag}, \ref{fig:TAW_geo_DMD}, and \ref{fig:MRW_geomag} are publicly available. 
%%The simulation data used in this study are available upon request.

%\subsection*{Code availability}
%%\bmhead{Code availability}
%The codes used in this study are available upon request.

\quad

%\clearpage
\appendix
%\begin{appendices}



\section{Normal mode calculations} \label{sec:normal_modes}

%An appendix contains supplementary information that is not an essential part of the text itself but which may be helpful in providing a more comprehensive understanding of the research problem or it is information that is too cumbersome to be included in the body of the paper.

Normal mode solutions of torsional waves (\ref{eq:TAW}) and
 of magnetic Rossby waves (\ref{eq:ode_axis-azimuth-field}) are computed individually, 
 whilst considering a basic state in a spherical shell outside the tangent cylinder.

\begin{figure}[hb]
\centering
\begin{tabular}{ll}
 \hspace{5mm}(a) & \hspace{10mm}(b) \\
 \includegraphics[width=0.46\linewidth]{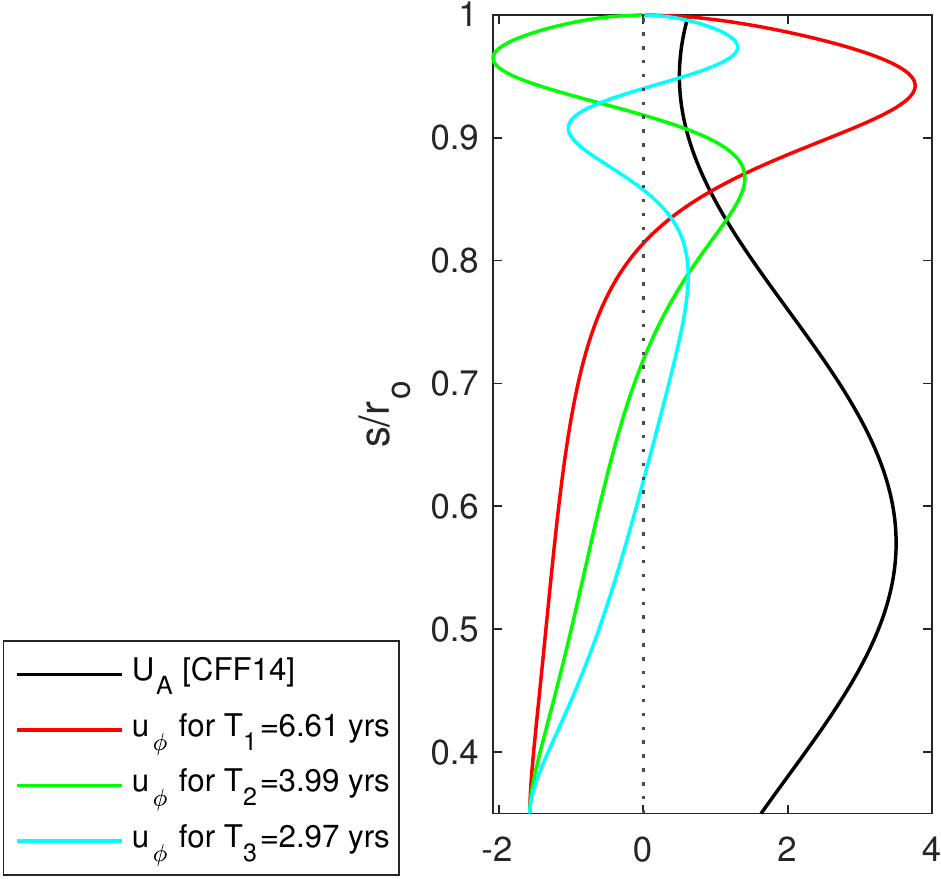} & 
 \hspace{5mm}\includegraphics[width=0.45\linewidth]{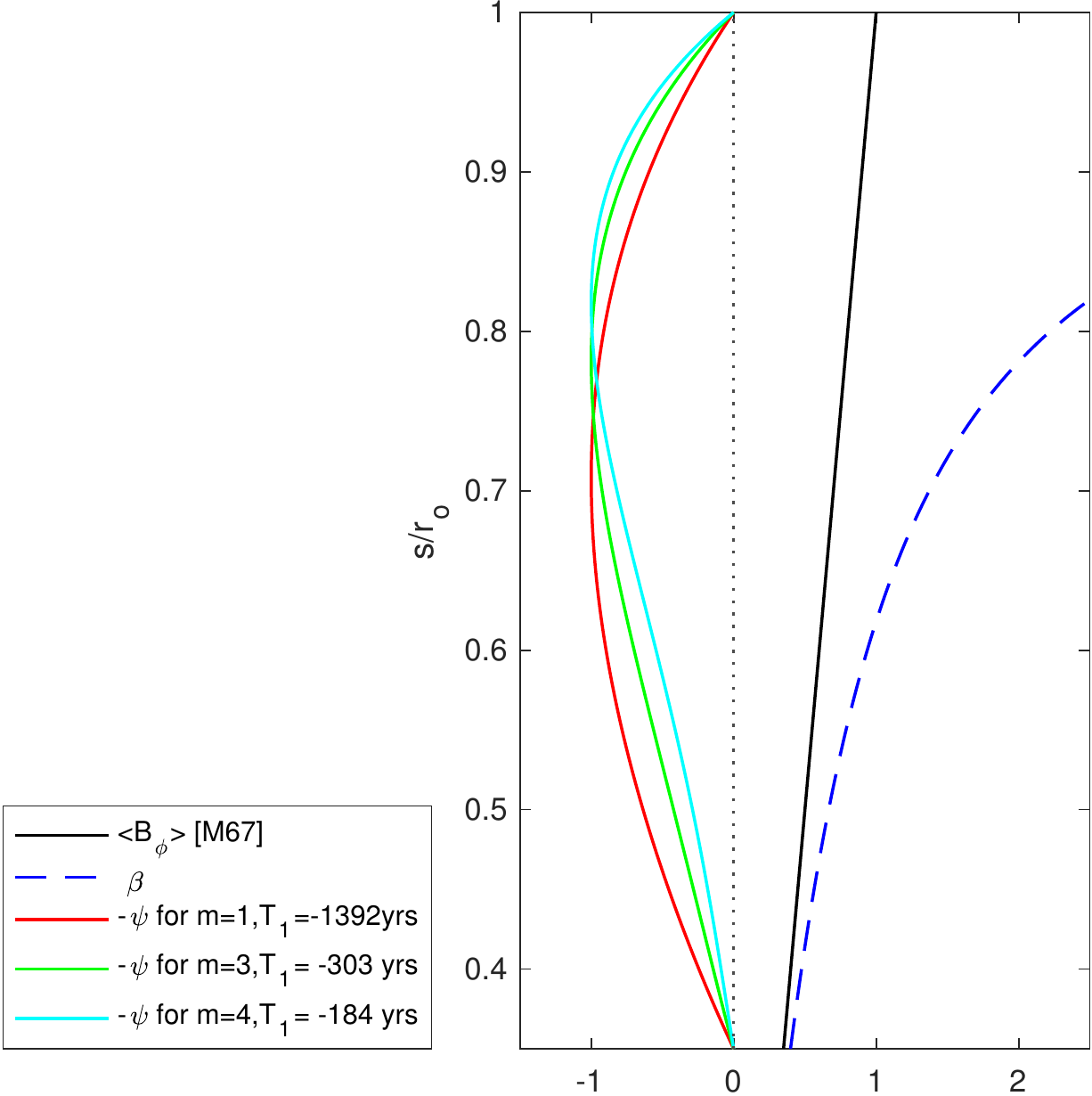}
\end{tabular}
 \caption{(a) Normal modes of torsional oscillations (\ref{eq:TAW}).
 The black solid curve shows the normalised profile of the given background field 
 $U_\text{A}$ \citep{GJCF10,CFF14}.
 Other curves show the eigenfunctions $\hat{u}_\phi$ of the 1st (red), 2nd (green), and 3rd (cyan) normal modes.
 The eigenvalue of the $i$-th normal mode is listed in the legend,
 in which the period $T_i$ is represented in years for a maximal background poloidal field 3.9 mT. 
 (b) Normal modes of magnetic Rossby waves (\ref{eq:ode_axis-azimuth-field}).
 The black solid and blue dashed curves show the normalised profiles
 of a background field $\langle \overline{\widetilde{B_\phi}} \rangle$ \citep{M67} and beta parameter $\beta$, respectively.
 Other curves show the eigenfunctions $\hat{\psi}$ of the 1st normal mode for $m =$ 1 (red), 3 (green), and 4 (cyan).
 For visualisation their profiles are presented in the negative domain, as $-\hat{\psi}$.
 Periods $T_i$ of the $i$-th eigenvalue are represented in years,
 given a background toroidal field of maximal magnitude 13 mT. }
 \label{fig:TW_normal_modes}
\end{figure}

For normal mode solutions, e.g. $\langle \overline{u'_\phi} \rangle = \hat{u}_\phi (s) \exp{\mathrm{i} \omega t}$,
 the homogeneous part of (\ref{eq:TAW}) implies a second-order ODE with variable coefficients. 
This is an eigenvalue problem to determine the eigenfunction $\hat{u}_\phi$ with eigenvalue $\omega$,
 given a basic state of $U_\text{A}$ and $\langle \overline{\rho} \rangle$. 
Here we suppose a basic field, $U_\text{A} \propto (3/2) \cos{\{ \pi(3/2 - 50s/19r_\text{o}) \}} + 2$, 
 which was adopted by \citet{CFF14}
 to represent the 1d structure of the poloidal field part in the geodynamo \citep{GJCF10}.
The density $\overline{\rho}$ is assumed to be constant.
We use the Matlab routine bvp4c to solve the eigenvalue problem in $0.35 \le s/r_\text{o} < 1$.
The boundary conditions are $d\hat{u}_\phi/ds = 0$ at the inner boundary
 and ${\hat{u}_\phi + (1-s/r_\text{o}) d\hat{u}_\phi/ds} = 0$ at $s/r_\text{o} = 0.99999$:
 the latter is introduced to avoid the numerical issue for singularities \citep{HTJ20}. 
Figure~\ref{fig:TW_normal_modes}a depicts profiles of eigenfunctions $\hat{u}_\phi$,
 for which a normalising factor is imposed at the inner bound.
Their eigenvalues $\omega$ are listed in the legend in terms of dimensional periods $T = 2\pi/\omega$.
For calculating the dimensional values we use $\rho = 1.13 \times 10^4\,kg/m^3$,
 $r_\text{o} = r_\text{core} = 3.485 \times 10^6\,m$,
 and a factor of $1.12\times 10^{-3}\,T$ for $\overline{\widetilde{B_s}}$,
 implying the maximal strength of the assumed backgound field is about 3.9 mT.

Similarly, the eigenvalue problem (\ref{eq:ode_axis-azimuth-field}) for magnetic Rossby waves is solved for the eigenfunction $\hat{\psi}$ and the eigenvalue $\hat{\omega}$,
 given a basic state $\omega_\text{M}$, $\beta$, and azimuthal wavenumber $m$. 
We retain the constant density assumption,
 and assume a simple profile for the basic field, $\overline{\widetilde{B_\phi}} \propto s/r_\text{o}$ \citep{M67},
 since the structure of the toroidal field in the geodynamo is unknown. 
The beta parameter is given topographically (\ref{eq:beta_topo0}), i.e. $\beta \propto s/(1-s^2/r_\text{o}^2)$.
The inner boundary condition is now $\hat{\psi} = 0$,
 while the modified condition is again adopted at the outer boundary.
Figure~\ref{fig:TW_normal_modes}b demonstrates the first normal modes for $m = 1, 3,$ and $4$.  
The eigenvalues are presented in the legend,
 where we suppose the dimensional quantities above and
 additionally $\Omega = 7.29\times 10^{-5}\,s^{-1}$
 and a maximal strength $\overline{\widetilde{B_\phi}}$ of 13 mT. \\

%%%% Figure A1 %%%%%%%%%%%%%%%%%%%%%%%%%%%%%%%%%%%%%%%%%%%%%%%%%%%%%%%%%%%%%
\bigskip 
%%%%%%%%%%%%%%%%%%%%%%%%%%%%%%%%%%%%%%%%%%%%%%%%%%%%%%%%%%%%%%%%%

%\quad

\section{WKBJ solutions}   \label{sec:WKBJ}

In order to gain insight of nonaxisymmetric wave motion (\ref{eq:ode_axis-azimuth-field}),
 we suppose the coefficients slowly vary in $s$ and seek a WKBJ solution. 
Rewriting the ODE as  
\begin{equation}
 \frac{d^2 \hat{\psi}}{d s^2} 
  + \frac{1}{s}  \frac{d \hat{\psi}}{d s} 
  + \lambda^2 \hat{\psi} = 0  
  \qquad \textrm{where} \qquad 
 \lambda^2 = \frac{\hat{\omega} \beta m}{ (\hat{\omega}^2 - \omega_\text{M}^2 )s} - \frac{m^2}{s^2}\; , 
\end{equation}
we seek solutions in form of $\hat{\psi} = A(s) \exp{\mathrm{i} \theta(s)}$. 
The ODE is then split into real and imaginary parts:  
\begin{equation}
 \frac{d^2 A}{ds^2} - A \left( \frac{d \theta}{ds} \right)^2
 + \lambda^2 A = 0 \qquad \textrm{and} \qquad  
 \frac{d^2 \theta}{ds^2} + \frac{2}{A} \frac{dA}{ds} \frac{d\theta}{ds}
 + \frac{1}{s}
   \frac{d\theta}{ds} = 0 \; ,
\end{equation}
respectively.
The amplitude is assumed to be slowly varying so that the highest order term
 of the real part is small, compared with other terms. 
Substituting this into the imaginary part gives
\begin{equation}
 \frac{d^2 \theta}{ds^2}
 = - \left( \frac{2}{A} \frac{dA}{ds} + \frac{1}{s}  \right) \frac{d\theta}{ds} 
 = - 2s \left[   \left( \frac{d\theta}{ds} \right)^2 - \lambda^2   + \frac{1}{2s^2} \right] \frac{d \theta}{ds} \;.
\end{equation}
If $d^2 \theta/ds^2$ is also smaller than the other terms
 the equation is drastically simplified.
Wave-like solutions then exist when $\lambda^2 > 1/2s^2$, leaving 
\begin{equation}
 A = 
 \frac{A_0}{\sqrt{s}}
 \qquad \textrm{and} \qquad 
 \left( \frac{d \theta}{d s} \right)^2  = \lambda^2 - \frac{1}{2s^2} 
   \; . \label{eq:wkb-solutions}
\end{equation} 
The local radial wavenumber
 $d\theta/ds$ (we denote as $n$) is determined by $\hat{\omega}$, $\beta$, 
 $m$, $s$, and $\omega_\text{M}^2$.   
The solutions become evanescent when $\lambda^2 < 1/2s^2$. 
Also it implies that the radial velocity, $\langle u'_s \rangle$, varies with
 $s^{-3/2}$. 

%\end{appendices}

\quad

%\begin{thebibliography}{9}

%\clearpage
%%%%%%%%%%%%%%%%%%%%%%%%%%% references

\end{document}